\documentclass[reprint,superscriptaddress,amsmath,amssymb,aps,twocolumn]{revtex4-1}
\usepackage{graphicx}
\usepackage{bm}
\usepackage{amsmath,amssymb,amsfonts}
\usepackage{epsfig}
\usepackage{epstopdf}
\usepackage{dcolumn}
\usepackage{grffile}
\usepackage{verbatim}
\usepackage{mathrsfs}
\usepackage{appendix}
\usepackage{xr}
\usepackage{extarrows}
\usepackage[normalem]{ulem}
\usepackage[colorlinks=true,linkcolor=blue,citecolor=blue, urlcolor=blue]{hyperref}

\newcommand{\bfk}{\mathbf{k}}
\newcommand{\bfq}{\mathbf{q}}
\newcommand{\e}{\mathrm{e}}

\newcommand{\dd}{\mathrm{d}}
\newcommand{\Tr}{\mathrm{Tr}}
\newcommand{\HBCO}{\mathrm{HgBa}_2\mathrm{CuO}_{4+\delta}}

\begin{document}
\setlength{\abovedisplayskip}{3pt}     
\setlength{\belowdisplayskip}{3pt}
\graphicspath{{figures/}}

\title{Yamaji effect and quantum oscillation in Yang-Rice-Zhang model of underdoped cuprates}

\author{Yicheng Zhong}
\affiliation{Beijing National Laboratory for Condensed Matter Physics and Institute of Physics, Chinese Academy of Sciences, Beijing 100190, China}
\affiliation{School of Physical Sciences, University of Chinese Academy of Sciences, Beijing 100190, China}

 \author{Fu-Chun Zhang}
 \email{fuchun@ucas.ac.cn}
 \affiliation{Kavli Institute for Theoretical Sciences, University of Chinese Academy of Sciences,
 	Beijing, 100190, China}

\author{Kun Jiang}
\email{jiangkun@iphy.ac.cn}
\affiliation{Beijing National Laboratory for Condensed Matter Physics and Institute of Physics, Chinese Academy of Sciences, Beijing 100190, China}
\affiliation{School of Physical Sciences, University of Chinese Academy of Sciences, Beijing 100190, China}

\date{\today}

\begin{abstract}
Recent experiments have revealed signatures of small Fermi pockets in the pseudogap phase of cuprate superconductors, most notably the Yamaji effect observed in $\HBCO$. The Yang-Rice-Zhang (YRZ) model provides a successful phenomenological description of the pseudogap state and naturally predicts such small pockets. In this work, we use a microscopic framework to calculate angle-dependent magnetoresistance and quantum oscillation within the YRZ model. Our calculations simultaneously reproduce the experimentally observed Yamaji oscillations and the Shubnikov–de Haas oscillation corresponding to a pocket area of about $p/8$, with $p$ the hole density. By further testing the effect of Green’s-function zeros, we confirm that isolated zeros leave the oscillation period unchanged, whereas an extended zero segment suppresses and modifies the oscillation. Our findings demonstrate that the YRZ model captures essential features of the pseudogap regime and provides a general quantum approach that can be applied to more complex electronic structures.
\end{abstract}

\maketitle

\textit{Introduction} The underdoped cuprate superconductors exhibit an unconventional normal state dominated by strong electronic correlations \cite{Anderson_2004,keimer_review,Timusk_1999,rice2011phenomenological}. In the pseudogap regime, the low-energy spectral weight is partially suppressed in a momentum-selective manner, strongly affecting the antinodal regions \cite{Timusk_1999,Sebastian_2012,NMR_PhysRevLett.62.1193,NMR_PhysRevLett.63.1700,NMR_PhysRevLett.70.2012,optical_1996,optical_PhysRevLett.71.1645,Transport_PhysRevLett.70.3995}. 
One of the most striking manifestations of this regime is the appearance of Fermi arcs in angle-resolved photoemission spectroscopy—finite segments of spectral weight near the nodal direction, instead of a closed Fermi surface expected for a conventional metal \cite{Shen_PhysRevLett.76.4841,ding_1996,norman1998destruction,kyle_shen,HBYang_PhysRevLett.107.047003,Uchida_PhysRevB.79.140502,HBYang_nature}. These features reflect the profound Fermi-surface reconstruction that occurs as the cuprates evolve from a Mott insulator at zero doping toward overdoping.

A prominent framework for describing the pseudogap phase is the Yang–Rice–Zhang (YRZ) model~\cite{YRZ, Chen_2008, rice2011phenomenological,Yang_2009, Rice2017PhysRevB.96.220502, Zhanglong}. In this approach, a phenomenological Green’s function is constructed from a renormalized $t-J$ Hamiltonian motivated by Anderson’s resonating–valence–bond (RVB) picture~\cite{anderson1987resonating, zhang1988effective, zhang1988renormalised}. The YRZ ansatz naturally captures the emergence of Fermi arcs through the formation of a Luttinger surface of zeros that truncates the Fermi surface into small pockets with highly anisotropic spectral weight. It also predicts a Lifshitz transition in the Fermi surface topology as the hole doping increases, providing a coherent phenomenological description of the evolution from the pseudogap state to a more conventional metallic state \cite{davis_stm,HBYang_nature,kanigel2006evolution,Luttingersurface,rice2011phenomenological}.

Recent angle-dependent magnetoresistance (ADMR) measurements on $\mathrm{HgBa_2CuO_{4+\delta}}$ (Hg1201)~\cite{chan2025observation} shed new insights into the Fermi surface of the pseudogap state. These experiments reveal an oscillatory modulation of the c-axis resistivity as a function of the magnetic-field tilt angle, namely the Yamaji effect \cite{yamaji1989angle,yagi1990semiclassical,kurihara1992microscopic,fang2022fermi}. The data suggest that the normal state contains four small Fermi-surface pockets, each with area $p/8$ (in units of the Brillouin-zone area), where $p$ is the hole doping level. A semiclassical analysis ~\cite{zhao2025yamaji} has shown that such pocket areas are consistent with a fractionalized Fermi liquid, which effectively reduces the pocket area from $p/4$ to $p/8$ due to the presence of emergent gauge fields and fractionalized excitations~\cite{sachdev2003,sachdev2010,chatterjee2016,zhangyahui2020}.

In this work, we offer an alternative interpretation of the Yamaji effect grounded in the YRZ framework. 
The YRZ model predicts a small Fermi pocket with area $p/8$ \cite{YRZ,rice2011phenomenological}.
Because the YRZ ansatz is formulated at the level of a phenomenological Green’s function with zeros, we employ a microscopic calculation of ADMR using the Landau-level basis \cite{abrikosov1969galvanomagnetic,kurihara1992microscopic}. This avoids relying on the semiclassical Boltzmann description, which breaks down in regimes lacking well-defined quasiparticles. We extend this approach to analyze quantum oscillations \cite{Chen_2008,Zhanglong}, providing a different probe of the Fermi-pocket area. We further investigate the oscillatory phenomena in the presence of Green’s-function zeros associated with Fermi arcs. Our study delivers a unified and comprehensive picture of magnetic oscillations in the pseudogap state.

\begin{figure}[thb]
    \centering
    \includegraphics[width=\linewidth]{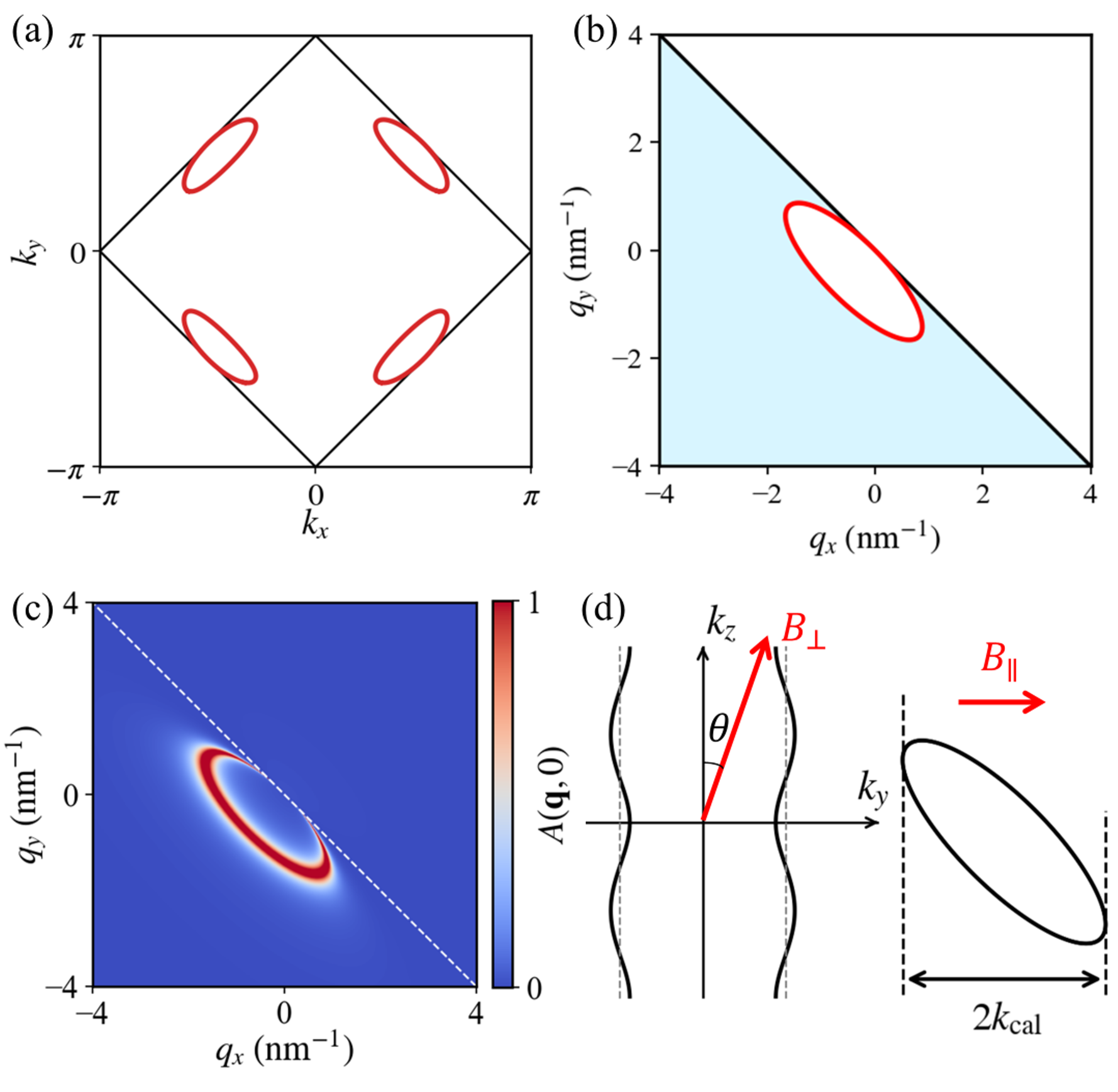}
    \caption{(a) and (b) shows the Fermi pocket of YRZ model by (a) tight binding approximation in Eq.~\ref{eq:GF} and (b) $k\cdot p$ approximation. The black line indicates the Luttinger surface of zeros and the red line represents the infinities of Green's function. In panel (b) within the blue area $G(\bfk,0)>0$.
    (c) Spectral function $A(\mathbf{q},0)$ around one pocket in $k\cdot p$ expanded YRZ model.
    (d) Schematic of Fermi surface in a quasi-2D system. Left: yz cross section with the out-of-plane magnetic field $B_{\perp}$ tilted from c-axis by polar angle $\theta$. Right: xy cross section with caliper momentum $k_{\mathrm{cal}}$ along the in-plane magnetic field direction $B_{\parallel}$ marked by an arrow.}
    \label{fig:fs}
    \vspace{-12pt}
\end{figure}

\textit{YRZ Ansatz}
The essential assumption of the YRZ model is the self-energy $\Sigma(\mathbf{k},\omega)$ of the coherent part of Green's function, which takes the form $\frac{\Delta_R^2}{\omega+\xi_0(\mathbf{k})}$, inspired by the form of the Green’s function in a doped spin liquid \cite{KRT_PhysRevLett.96.086407,YRZ,Ng_PhysRevB.71.172509,Kotliar_PhysRevB.74.125110}. Then, the Green's function can be written as
\begin{equation}
G_\mathrm{YRZ}(\mathbf{k}, \omega)
= \frac{g_t}{\omega - \xi(\mathbf{k}) - 
\frac{\Delta_R^2(\mathbf{k})}{\omega + \xi_0(\mathbf{k})}} ,
\end{equation}
where $\mathbf{k} = (k_x, k_y)$ in a two-dimensional Brillouin Zone, $g_t$ is renormalization factor and $\xi_0(\mathbf{k})= -2t(\cos k_x + \cos k_y)$, 
$\Delta_R(\mathbf{k}) = \Delta_0(\cos k_x - \cos k_y)$,
$\xi(\mathbf{k}) = \xi_0(\mathbf{k}) - 4t'\cos k_x \cos k_y - 2t''(\cos 2k_x + \cos 2k_y) - \mu$.
The parameters $g_t$, $t/ t'/t''$, and $\Delta_0$ are all doping-dependent within the renormalized mean-field framework \cite{YRZ}. A key feature of the YRZ ansatz is that the zero-frequency self-energy $\Sigma(\mathbf{k},0)$ diverges along $\xi_0(\mathbf{k})=0$, producing a line of zeros in the Green’s function $G_\mathrm{YRZ}(\mathbf{k}, 0)$. This Luttinger surface of zeros coincides with the antiferromagnetic Brillouin zone boundary, as shown in Fig.~\ref{fig:fs}(a).

It is also instructive to examine the loci where $G(\mathbf{k}, 0)$ diverges, which define the Fermi surface in a conventional metal. In the YRZ model, the resulting Luttinger surface is shown in Fig.~\ref{fig:fs}(a): four small Fermi pockets appear, each tangent to the line of zeros. A notable feature is the strong anisotropy of the quasiparticle spectral weight around these pockets—it is heavily suppressed along the outer edge near the zero surface and vanishes entirely at $(\pm\pi/2, \pm\pi/2)$. This naturally accounts for the Fermi arc phenomenology observed in the pseudogap regime \cite{YRZ}. With increasing hole doping, the small pockets continuously evolve into a large Fermi surface, demonstrating how the YRZ ansatz interpolates between a doped RVB spin liquid at low doping and a conventional Landau Fermi liquid at higher doping \cite{YRZ}.

Throughout this work, we concentrate on the underdoped regime (small $p$). As noted in the introduction \cite{Chen_2008}, each Fermi pocket in Fig.~\ref{fig:fs}(a) encloses an area $S= p/8$, rather than $p/4$. 
For analytical simplicity, we adopt a $k\cdot p$ expansion around a single pocket located near  $\mathbf{k}_0=(\pi/2, \pi/2)$ with $\mathbf{k}=\mathbf{k}_0+\mathbf{q}$, as illustrated in Fig.~\ref{fig:fs}(b). Within this approximation, the Green function denominator reduces to 
$\omega- (\gamma_\bfq-\mu + \Delta_\bfq^\dagger (\omega+\gamma_\bfq)^{-1}\Delta_\bfq)$ with $\gamma_\bfq=2ta_0(q_x+q_y)$, $\Delta_\bfq=\Delta_0a_0(q_x-q_y)$, where the pseudogap magnitude defined in the YRZ model, and $a_0$ is the in-plane lattice constant. We omit $t'/t''$ for simplicity.
Importantly, this simplified treatment leaves the pocket and its key properties unchanged. All of the following analyses become asymptotically exact once the four pockets are well separated. We also note that the specific parameter choices entering $\gamma$ and $\Delta$ do not affect our conclusions regarding magnetic oscillations, provided the Fermi pocket area remains intact.

\textit{Yamaji Effect}
The Yamaji effect originates from the quasi-two-dimensional (quasi-2D) electronic structure of layered materials. Because of the weak interlayer coupling, the Fermi surface evolves from a perfect cylinder in the purely 2D limit to a weakly warped cylinder in the quasi-2D case, as shown in Fig.~\ref{fig:fs}(d). This warping causes the cross-sectional area of the Fermi surface perpendicular to the magnetic field to vary as the field is tilted. Consequently, the interlayer magnetoresistance oscillates when the magnetic field deviates from the crystallographic c-axis. At specific magic angles, known as the Yamaji angles, the cyclotron-orbit cross sections become degenerate, leading to constructive interference and a pronounced enhancement of the oscillation amplitude. These maxima occur when the tilt angle $\theta$ satisfies $c_0k_\mathrm{cal}\tan\theta=\pi(N-\frac{1}{4})$
as derived in Refs.~\cite{yagi1990semiclassical, kurihara1992microscopic}, where $c_0$ is the c-axis lattice constant, $k_\mathrm{cal}$ is the caliper momentum along the the in-plane magnetic field direction and $N$ is the integer for Yamaji peak order. Therefore, the geometry of the Fermi surface can be obtained by tuning the azimuthal angle $\phi$ of the in-plane magnetic field.

To evaluate this effect, we adopt the magnetic-field geometry shown in Fig.~\ref{fig:fs}(d), where the field $\mathbf{B}$ is tilted by an angle $\theta$ within the yz-plane. We introduce the gauge potential $\mathbf{A} = (0, Bx\cos\theta, -Bx\sin\theta)$ into the Green's function via minimal coupling. Under this choice of gauge, the guiding-center coordinate $X_0$ and the momentum $k_z$ remain good quantum numbers and it is natural to describe the system with Landau level basis $|n,X_0,k_z\rangle$, with $n$ the index of Landau levels. As the Fermi pocket is approximately an ellipse, we use Landau levels of anisotropic quadratic dispersion to reduce numerical errors. As for the azimuthal angle $\phi$, we can rotate the xy-plane instead of magnetic field for simplicity. Then, the retarded Green's function can be written as
\begin{equation}
    G(\bfk,\omega)=\frac{1}{\omega-(\gamma_\bfk-\mu)-\Delta_\bfk^\dagger (\omega+\gamma_\bfk)^{-1}\Delta_\bfk-H_z(\bfk)+\mathrm{i}\Gamma},
    \label{eq:GF}
\end{equation}
where the interlayer coupling term $H_z=-2t_z\mathrm{cos}((k_z-eA_z/\hbar)c_0)$ with interlayer hopping magnitude $t_z$, and the imaginary term $\mathrm{i}\Gamma$ represents dissipation. We also set $g_t=1$, which only influences resistivity amplitudes.

Then, the interlayer conductivity can be calculated with Kubo formula, as 
\begin{equation}
    \sigma_{zz}=\frac{e^2\hbar}{2\pi}
    \sum_{k_z, X_0}
    \int\dd E\,
    \Big(-\frac{\partial f}{\partial E}\Big)\,
    \Tr\!\big[v_z A(E)v_z A(E)\big],
    \label{eq:Kubo}
\end{equation}
where z-direction velocity $v_z=\frac{\partial H_z}{\partial p_z}$ and spectral function $A(E)$ are matrix under Landau level basis. 
Only the first order current-current correlation is taken into account. 
The c-axis resistivity can be obtained as $\rho_{zz}\approx1/\sigma_{zz}$. 

Having established the quantum formulation, it is important to demonstrate how this approach connects with the semiclassical one. Considering a 2D free-electron system with interlayer hopping, matrix elements of $H_z$ can be written on Landau basis as $H_{z,nn}= -2t_z\cos\!\big(c_0k_z+\kappa X_0\big)\e^{-\alpha/2}L_n(\alpha)$, where $\kappa ={eBc_0\sin\theta }/{\hbar}$, $\alpha=\tfrac{\hbar\kappa^2}{2eB\cos\theta}$ and $L_n$ is the n-th-order Laguerre function. In the semi-classical limit, the magnetic field is small enough, so the Landau level $n_F$ corresponding to the Fermi surface is very large. We can use the asymptotic relation of large-$n$ Laguerre function and Bessel function so that $\e^{-\alpha}L_{n_F}^2(\alpha)\approx\ J_0^2\!\big(k_\mathrm{cal} c_0\tan\theta\big)$, where $J_0$ is the zeroth-order Bessel function. Taking the zero points of the Bessel function, we can come back to the Yamaji condition: $c_0k_\mathrm{cal}\tan\theta=\pi(N-\frac{1}{4})$ \cite{kurihara1992microscopic}.

\begin{figure}[hb]
    \centering
    \includegraphics[width=\linewidth]{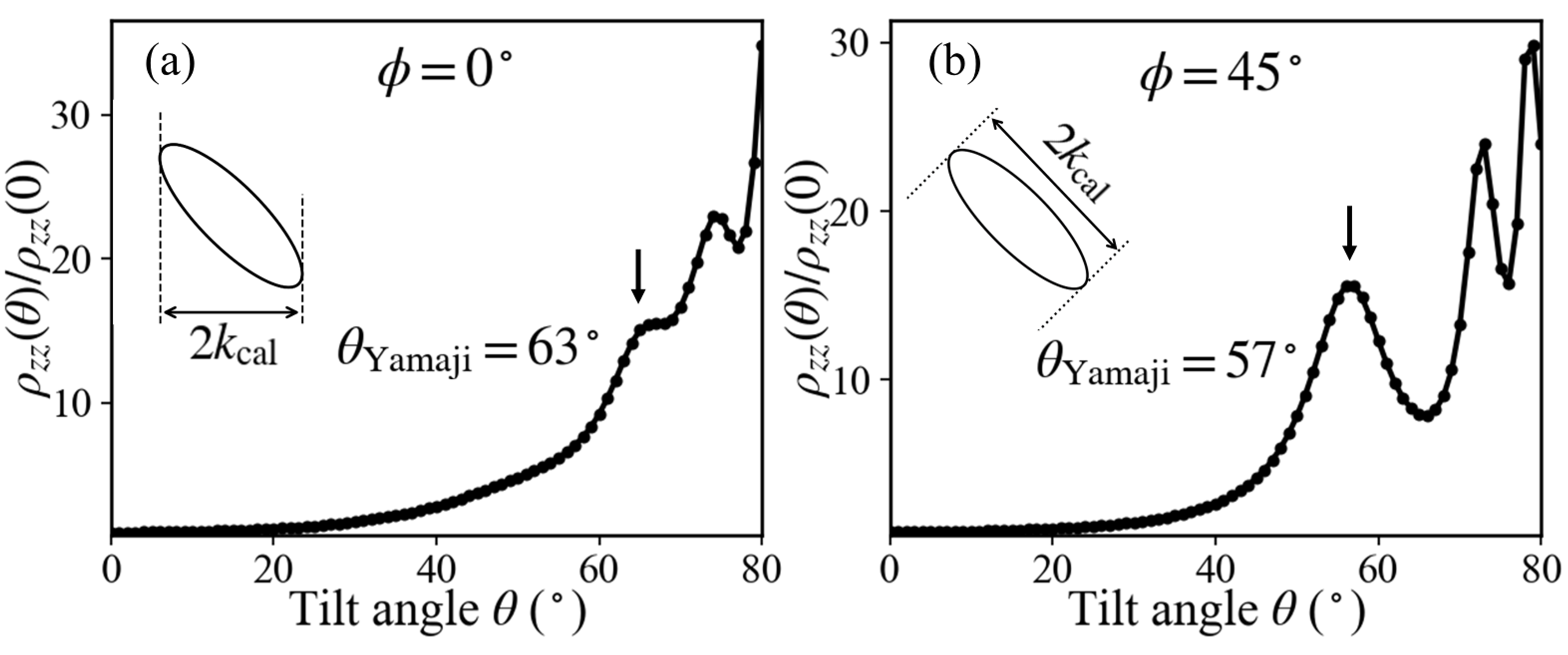}
    \caption{Calculated interlayer resistivity $\rho_{zz}(\theta)/\rho_{zz}(0)$ as a function of the polar angle $\theta$. Panel (a) and (b) are results with azimuthal angle $\phi=0^\circ$ and $\phi=45^\circ$. The first Yamaji peak is marked with an arrow. The inset schemtically shows the caliper radius corresponding to $\phi$ in each case.}
    \label{fig:yamaji}
\end{figure}

With this information, we can calculate the resistivity as a function of the polar angle $\theta$ for different azimuthal angle $\phi$, as shown in Fig.~\ref{fig:yamaji}. The lattice constants are taken as $a_0=3.88~\text{\AA}$ and $c_0=9.50~\text{\AA}$. For the hopping parameters, we use $t=0.43$eV and an interlayer hopping amplitude $t_{z}=0.005$eV based on density functional calculation \cite{Kent_PhysRevB.78.035132}. 
To capture the essential features of the Fermi pockets at hole doping $p=0.1$ in Hg1201, we set $\Delta_0 = 0.28$eV to open the pseudogap and use $\mu = -0.524$eV, obtained from the Luttinger sum rule. The magnetic field is chosen as $B = 20~\mathrm{T}$; the Yamaji peak, however, persists over a broad range of the field from below $10~\mathrm{T}$ to above $100~\mathrm{T}$. To compute the resistivity, we include an imaginary self-energy $i\Gamma$ and set $\Gamma=0.03$eV, which controls the peak resolution in ADMR. Fig.~\ref{fig:yamaji}(a) and (b) display results for interlayer resistance with azimuthal angle $\phi=0^\circ$ and $\phi=45^\circ$ respectively.

The calculated ADMR curves show good agreement with experiment. As for $\phi=0^\circ$, the first Yamaji angle appears near $63^\circ$, close to the experimental value of $63.5^\circ$\cite{chan2025observation}. When $\phi=45^\circ$, the first Yamaji peak is around $57^\circ$, slightly larger than the measured value $56.5^\circ$\cite{chan2025observation}.
In both case there is a second Yamaji peak which is near $75^\circ$ and $72^\circ$ respectively, qualitatively corresponding to the experimentally observed upturn in resistivity at similar angles, though more precise measurements are needed to confirm this feature. From the Yamaji angle we can determine $k_\mathrm{cal}(\phi=0^\circ)=1.29\mathrm{nm}^{-1}$ and $k_\mathrm{cal}(\phi=45^\circ)=1.64\mathrm{nm}^{-1}$, and a Fermi pocket area around 1.5\% of the BZ.

\textit{Quantum Oscillation}
Interestingly, our method is not restricted to capturing the Yamaji effect; at sufficiently large magnetic fields, we also observe clear Shubnikov–de Haas (SdH) oscillations superimposed on the angular magnetoresistance signal. Such quantum oscillations are a hallmark of closed Fermi-surface orbits and have been reported and discussed in underdoped cuprates~\cite{doiron2007quantum,barivsic2013universal,Chen_2008,allais2014connecting,bonetti2024quantum,Zhanglong}. Together with the de Haas–van Alphen effect, SdH oscillations provide a standard and powerful probe of closed Fermi surfaces \cite{ashcroft1976solid,abrikosov2017fundamentals}. Their physical origin lies in Landau quantization: as the magnetic field is varied, discrete Landau levels periodically cross the Fermi energy, producing oscillations in measurable quantities. These oscillations are periodic in $1/B$, and the corresponding Fermi-surface area $S$ is determined from the oscillation period via $S=\tfrac{2\pi e}{\hbar\delta(1/B)}$ with $\delta(1/B)$ denoting the period.
Using the same framework as in the ADMR analysis, we compute the c-axis resistivity from Eq.~\ref{eq:Kubo} with the magnetic field applied strictly along the c-axis and varied between 50\text{ T} and 200\text{ T}. All remaining parameters are identical to those used in our ADMR calculations.

\begin{figure}[hb]
    \centering
    \includegraphics[width=\linewidth]{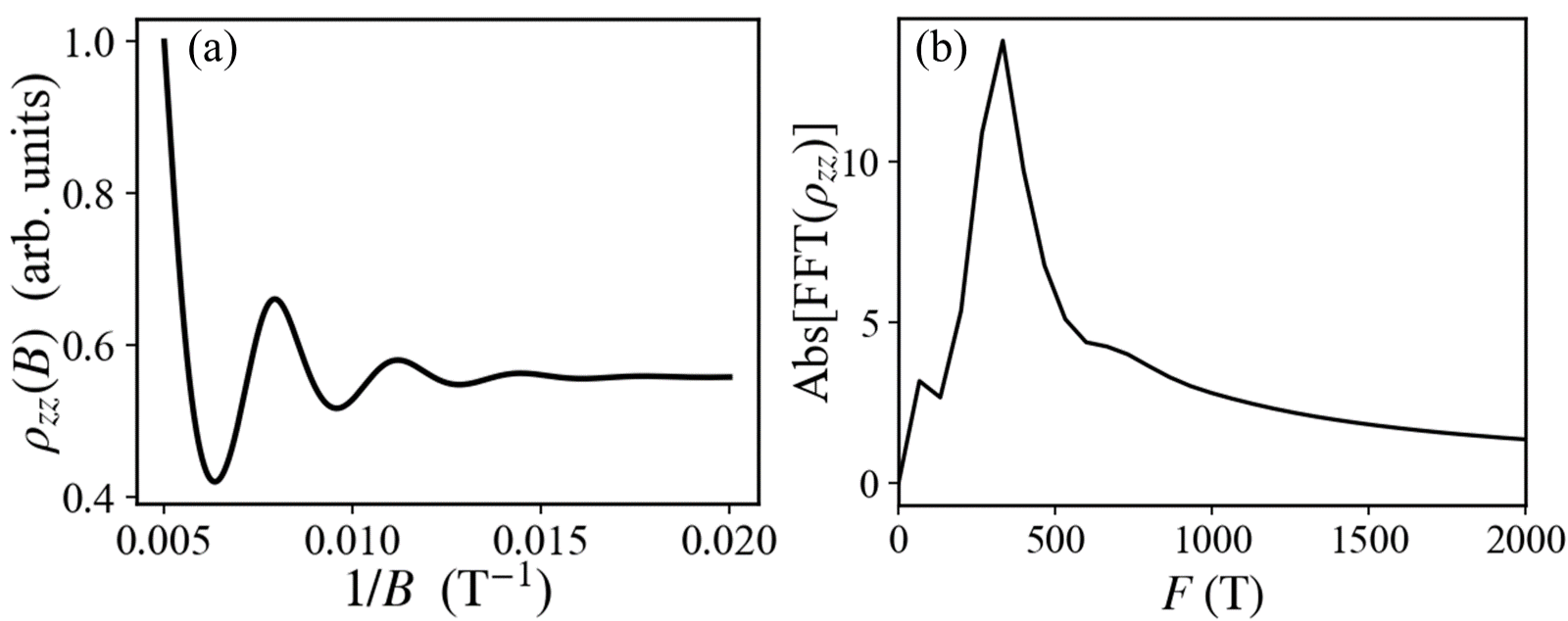}
    \caption{(a) Calculated interlayer resistivity $\rho_{zz}$ as a function of the inverse magnetic field $B^{-1}$. (b) Absolute value of Fourier transformed interlayer resistivity}
    \label{fig:sdh}
\end{figure}

The calculated interlayer resistance is shown in Fig.~\ref{fig:sdh}. Clear oscillations in $\rho_{zz}$ appear as a function of $1/B$, but their amplitude decays rapidly with increasing $1/B$, making SdH oscillations difficult to detect at low magnetic fields. A Fourier-transform analysis yields a frequency $F = 332 \mathrm{T}$ and an oscillation period $\delta(1/B) = 0.003 \mathrm{T^{-1}}$. The range of magnetic field is larger than the Yamaji value and partly beyond experimentally accessible fields. Such high fields are required because the Fermi surface pocket is very small, and impurity scattering in realistic samples further damps the oscillation amplitude. The extracted Fermi-pocket area is $3.17 \mathrm{nm^{-2}}$, or 1.21\% of the first Brillouin zone, slightly smaller than the 1.25\% predicted by the YRZ model. For completeness, we also compute the de Haas–van Alphen signal in Appendix C, which leads to the same conclusion.

\textit{Zeroes and Arcs}
Based on the above discussion, we have demonstrated that the YRZ model naturally exhibits both the Yamaji effect and magnetic quantum oscillations. In particular, we show that the zero of the Green’s function located within each Fermi pocket does not alter the cyclotron motion. This raises a broader question: how do additional Green’s function zeros modify the oscillatory response? In particular, what happens when a conventional closed Fermi surface evolves into a structure composed of both poles and zeros distributed along the Fermi points?

\begin{figure}[thb]
    \centering
    \includegraphics[width=\linewidth]{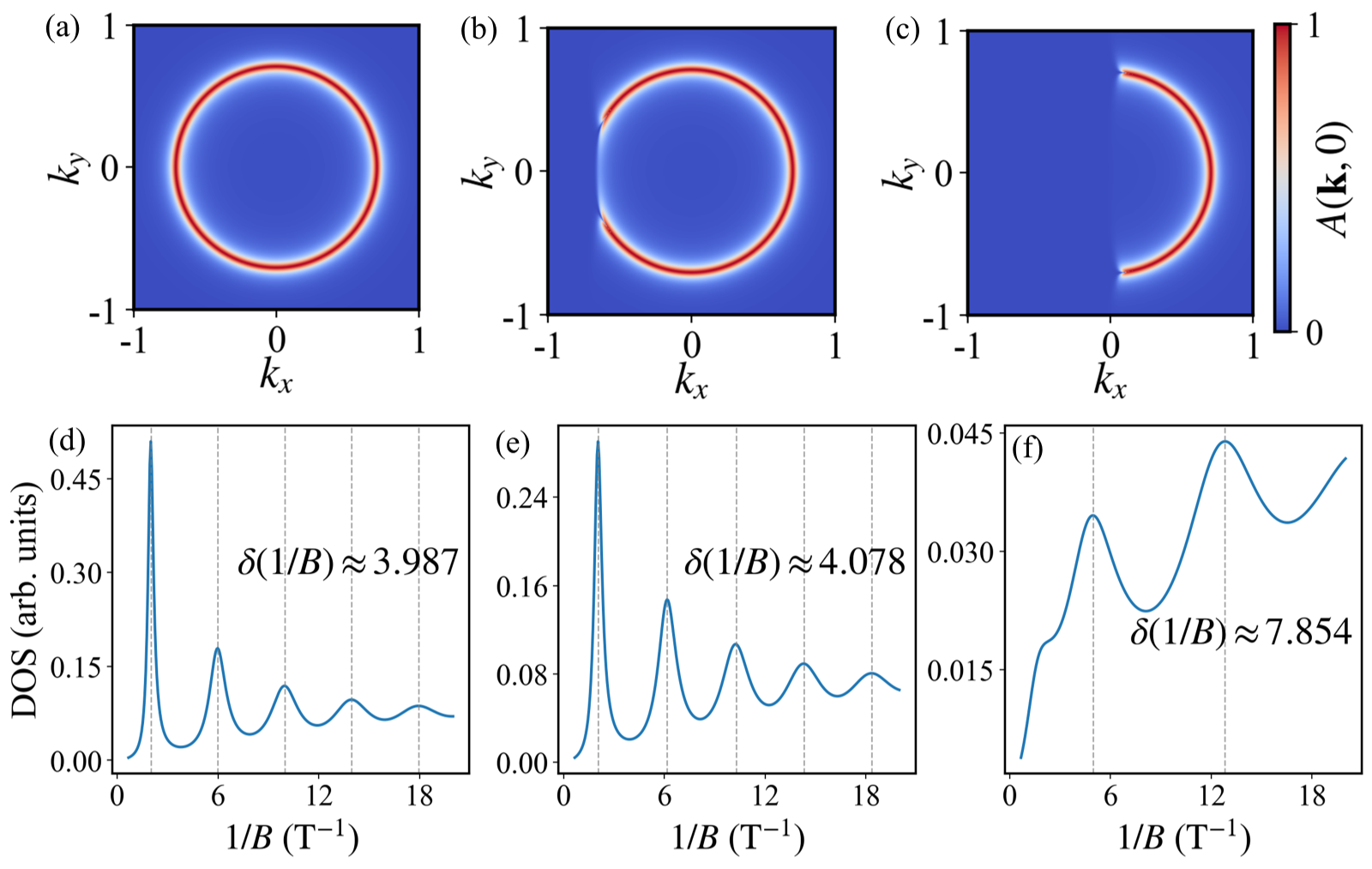}
    \caption{Illustration of role of Green's function zeroes and poles to the quantum oscillation. Spectral function $A(\mathbf{k}, \omega = 0)$ (panels (a)–(c)) with red lines as pole positions, and quantum oscillation in density of states (panels (d)–(f)) in Green's function Eq.\ref{eq:toy}.  
    Left panel: free electron case \(V_0 = 0\), in which there are no zeroes.  
    Middle panel: a small segment of zeroes with parameters $V_0 = 0.6,\; a = k_F =1/\sqrt{2},\; \mu = 0.5,\; \eta = 0.05$.
    Right panel: only half circle of poles with parameters as same as in the middle panel except $a = 0$.
}
    \label{fig:demo}
\end{figure}

To clarify the role of Green’s function zeros in the simplest possible setting, we construct a toy model based on a YRZ-type Green’s function for a single parabolic band:
\begin{equation}
    G_0(\bfk,\omega)=\frac{1}{\omega-\varepsilon(\bfk)-\frac{V(\bfk)^2}{\omega+\varepsilon(\bfk)}+i\Gamma},
    \label{eq:toy}
\end{equation}
where $\varepsilon(\bfk)=k_x^2+k_y^2-\mu$. We introduce a YRZ-like pseudogap self-energy $V(\bfk)=V_0\big(1-\frac{1}{2}(1+\tanh((k_x-a)/\eta)\big)$.
As $\eta \rightarrow 0$, $V(\bfk)$ approaches a Heaviside step function that eliminates the spectral weight for $k_x<a$. When $V_0 \to 0$, the expression reduces to the ordinary Green’s function of a 2D parabolic band. The resulting spectral function at the Fermi level, $A(\mathbf{k},\omega=0)$, is shown in Fig.~\ref{fig:demo}(a). For finite $V_0$, tuning the parameter $a$ selectively suppresses part of the spectral weight and produces zeros along the Fermi contour. For instance, choosing $a=k_F=1/\sqrt{2}$, $\mu=0.5$, $\eta=0.05$, and $V_0=0.6$, Fig.~\ref{fig:demo}(b) removes a small segment on the left side of the Fermi surface, generating a small line of zeros. Setting $a=0$ produces an even more dramatic effect: the entire left half of the Fermi contour is suppressed, giving rise to a Fermi arc terminated by zeros, as shown in Fig.~\ref{fig:demo}(c). In a conventional Fermi surface, the spectral weight exhibits a sharp boundary between occupied and unoccupied states. By contrast, zeros of the Green’s function correspond to states with mixed electron–hole character, giving rise to a strongly suppressed spectral weight at the Fermi level (see Appendix D for more discussion).

Next, we evaluate the density of states (DOS) as a function of $1/B$, with the corresponding results shown in Fig.~\ref{fig:demo}(d)–(f). The DOS is computed directly from the Green’s function via
\begin{equation}
    D(B) = -\frac{1}{\pi}\mathrm{ImTr}\,G_0(\bfk - e\mathbf{A},0).
\end{equation}
Comparing panel (d) and (e), we find that introducing only a few Green’s-function zeros produces an oscillation period nearly identical to the $V(k)=0$ case. In contrast, when the spectral weight is removed from half of the Fermi contour, the oscillation period doubles, corresponding to the area of a half-circle. In this regime, the DOS amplitude is also strongly suppressed due to the extensive loss of spectral weight. In addition, the oscillation period is roughly inverse proportional to the unsuppressed area. These results indicate that quantum oscillations are sensitive only when a finite segment of the Fermi contour is replaced by Green’s-function zeros. Consequently, in the YRZ model, the observed oscillation frequency genuinely reflects the area of the small Fermi pocket.

In summary, we develop a microscopic framework to compute ADMR and quantum oscillations within the YRZ description of the cuprate pseudogap. By implementing the YRZ Green’s function in the Landau-level basis and using a minimal $k\cdot p$ treatment of the small Fermi pocket, we reproduce both the Yamaji oscillations and the quantum-oscillation signals corresponding to a pocket area around $p/8$~\cite{YRZ,zhao2025yamaji,Kotliar_PhysRevB.74.125110}.
Although ADMR measurements in Hg1201 report such a $p/8$ pocket above $T_c$, we notice that the situation remains controversial at lower temperatures \cite{barivsic2013universal,chan_PNAS,Sebastian_2012}, where conventional quantum oscillations may be complicated by charge-density-wave reconstruction—particularly in YBa$_2$Cu$_3$O$_{6+\delta}$ \cite{YBCO}. Further experimental work will be essential for clarifying the true Fermi-surface topology of the pseudogap state.

We further show that the quantum-oscillation signal is insensitive to the zero of the YRZ Green’s function and, importantly, that oscillations originating from a closed Fermi pocket are fundamentally different from those associated with a Fermi arc terminated by an extended line of zeros. 
While our analysis focuses on the YRZ model, the framework is general and can be extended to explore quantum oscillations in other correlated systems. We hope that this framework and its results offer new insight into the electronic structure of the pseudogap state.

\textit{Acknowledgement}
We thank Subir Sachdev to bring the recent experiment Yamaji effect to our attention. We acknowledge the support by the National Natural Science Foundation of China (Grant NSFC-12574150, NSFC-12174428, NSFC-12494594), the Ministry of Science and Technology (Grant No. 2022YFA1403900), the Chinese Academy of Sciences Project for Young Scientists in Basic Research (2022YSBR-048), the Innovation program for Quantum Science and Technology (Grant No. 2021ZD0302500), and Chinese Academy of Sciences under contract No. JZHKYPT-2021-08.

\appendix
\newpage

\section{Connection of semi-classical and quantum approach to Yamaji effect}
To clarify how the microscopic calculation used in this work connects to the  semi-classical interpretation of the Yamaji effect, we consider the simplest case of a two-dimensional free-electron gas with a weak interlayer hopping. The energy dispersion is written as $\varepsilon(\bfk)=(k_x^2+k_y^2)/2m-2t_z\cos(c_0k_z)$ with $m$ the effective mass of the electron. When a tilted magnetic field $\mathbf{B}=(0,B_\parallel,B_\perp)$ is applied, the 2D energy becomes Landau levels with spacing $\tfrac{\hbar eB_\perp}{m}$. From Eq.~\ref{eq:Kubo} we find that the interlayer hopping term contributes to the c-axis conductivity mainly with $v_z$, and the interlayer hopping term can be omitted in spectral function $A(E)$. The diagonal matrix element of $v_z$ is
\begin{equation}
v_{z,nn}= 2c_0t_z\sin\!\big(c_0k_z+\kappa X_0\big)\e^{-\alpha/2}L_n(\alpha),
\end{equation}
where $\kappa={eBc_0\sin\theta }/{\hbar}$, $\alpha=\tfrac{\hbar\kappa^2}{2eB\cos\theta}$ and $L_n$ is the n-th order Laguerre function. As diagonal terms dominate in matrix $v_z$, we omit the off-diagonal elements here. In the zero temperature limit, the Kubo formula in Eq.~\ref{eq:Kubo} can be simplified to 
\begin{equation}
    \sigma_{zz}\simeq
    \frac{e^2 c_0^2 t_z^2}{\pi^2\hbar\,\Gamma}\,\e^{-\alpha}L_{n_F}^2(\alpha)\sum_{k_z,X_0}\sin^2\!\big(c_0k_z+\kappa X_0\big),
\end{equation}
from which we can see that the minimal of conductivity correspond to zero of $L_{n_F}(\alpha)^2$. With the asymptotic relation $\e^{-\alpha}L_{n_F}^2(\alpha)\approx\ J_0^2\!\big(k_\mathrm{cal} c_0\tan\theta\big)$ in the limit $n_F\gg1$, the conductivity can be written as 
\begin{equation}
    \sigma_{zz}(\theta)=\sigma_{zz}(0)J_0^2\!\big(k_\mathrm{cal} c_0\tan\theta\big),
\end{equation}
which is the same as semi-classical results~\cite{yagi1990semiclassical}. Higher-order terms originate from off-diagonal terms in $v_z$~\cite{kurihara1992microscopic}. 

\begin{figure}[thb]
    \centering
    \includegraphics[width=0.6\linewidth]{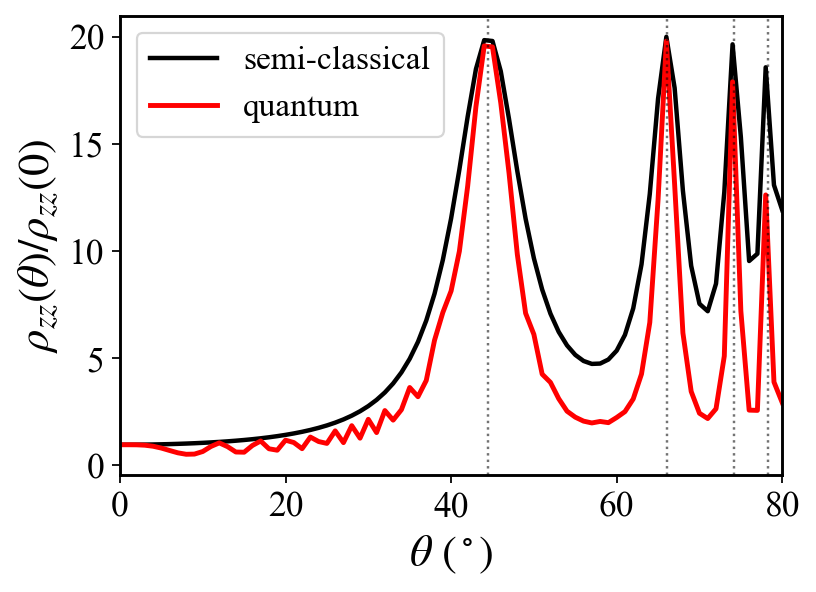}
    \caption{Calculated interlayer resistivity $\rho_{zz}(\theta)/\rho_{zz}(0)$ as a function of the polar angle $\theta$. The red curve represents results of quantum approcch, while the black curve represents results of semi-classical approach.}
    \label{fig:compare}
\end{figure}

To substantiate the discussion above, we directly compare the Yamaji oscillations obtained from the microscopic calculation with the semi-classical expression. As shown in Fig.~\ref{fig:compare}, the two approaches exhibit nearly identical angular dependence, confirming that the Yamaji effect arises naturally from the quantum formulation. In addition to the Yamaji maxima, SdH oscillations appear at small tilt angles, where the perpendicular field component remains large enough to produce Landau quantization. Their amplitude rapidly diminishes upon increasing $\theta$, consistent with the reduction of $B_\perp$, and becomes negligible once the system is effectively driven out of the quantum-oscillation regime.

\section{Calculation of ADMR for $\phi=45^\circ$}
When azimuthal angle $\phi\neq0$, the gauge potential will have a complex form and $X_0$ and $k_z$ are no longer good quantum numbers. This removal of degeneracy leads to far more matrix dimensions and complexity in calculation. However, as the Green's function has a simple form after $k\cdot p$ approximation, we can rotate the xy-plane for an azimuthal angle $\phi$ instead. Under this rotation, the in-plane momenta transform as $(k_x',k_y') = (k_x\cos\phi + k_y\sin\phi,\; -k_x\sin\phi + k_y\cos\phi)$. For $\phi=45^\circ$, this reduces to $k_x' = (k_x + k_y)/\sqrt{2}$ and $k_y' = (-k_x + k_y)/\sqrt{2}$. After the rotation, the $\gamma_\bfq$ and $\Delta_\bfq$ expressions in YRZ Green's function become $\gamma_\bfq' = 2t a_0\sqrt{2}\,q_x'$ and $\Delta_\bfq' = \Delta_0 a_0\sqrt{2}\,q_y'$. 

Working in the rotated frame restores $X_0$ and $k_z$ as good quantum numbers and allows us to reuse the same formalism as in the $\phi=0$ case. In our calculation, the Green's function is treated as a matrix with limited dimensions. To describe the anisotropic pocket precisely, we use Landau levels of anisotropic quadratic dispersion as the basis. The dispersion mentioned here has a form of $\varepsilon(\bfk)=k_x^2/2m_x+k_y^2/2m_y$ with $m_x/m_y=\Delta_0^2/4t^2$, since $\Delta_0/t$ is the aspect ratio of the ellipse pocket in Fig.~\ref{fig:fs}(b). After defining the rescaled canonical momentum $\pi_i = \sqrt{m_c/m_i}(k_i-eA_i)$ with $m_c=\sqrt{m_x m_y}$, the Hamiltonian becomes identical to the isotropic case with $H = (\pi_x^2 + \pi_y^2)/(2 m_c)$. The anisotropy is entirely encoded in the momentum matrix elements, which takes the form 
\begin{align}
    \pi_{x,nn'} = \sqrt{\frac{\hbar eBm_x}{2m_c}}(\sqrt{n'}\,\delta_{n,n'-1} + \sqrt{n'+1}\,\delta_{n,n'+1}),\\ \nonumber
    \pi_{y,nn'} = i\sqrt{\frac{\hbar eBm_y}{2m_c}}(\sqrt{n'}\,\delta_{n,n'-1} - \sqrt{n'+1}\,\delta_{n,n'+1}).
\end{align}
With these expressions, we can write the Green's function in Eq.~\ref{eq:GF} into matrix form and calculate ADMR with Eq.~\ref{eq:Kubo}.

\section{de Haas-van Alphen effect}
\begin{figure}[thb]
    \centering
    \includegraphics[width=\linewidth]{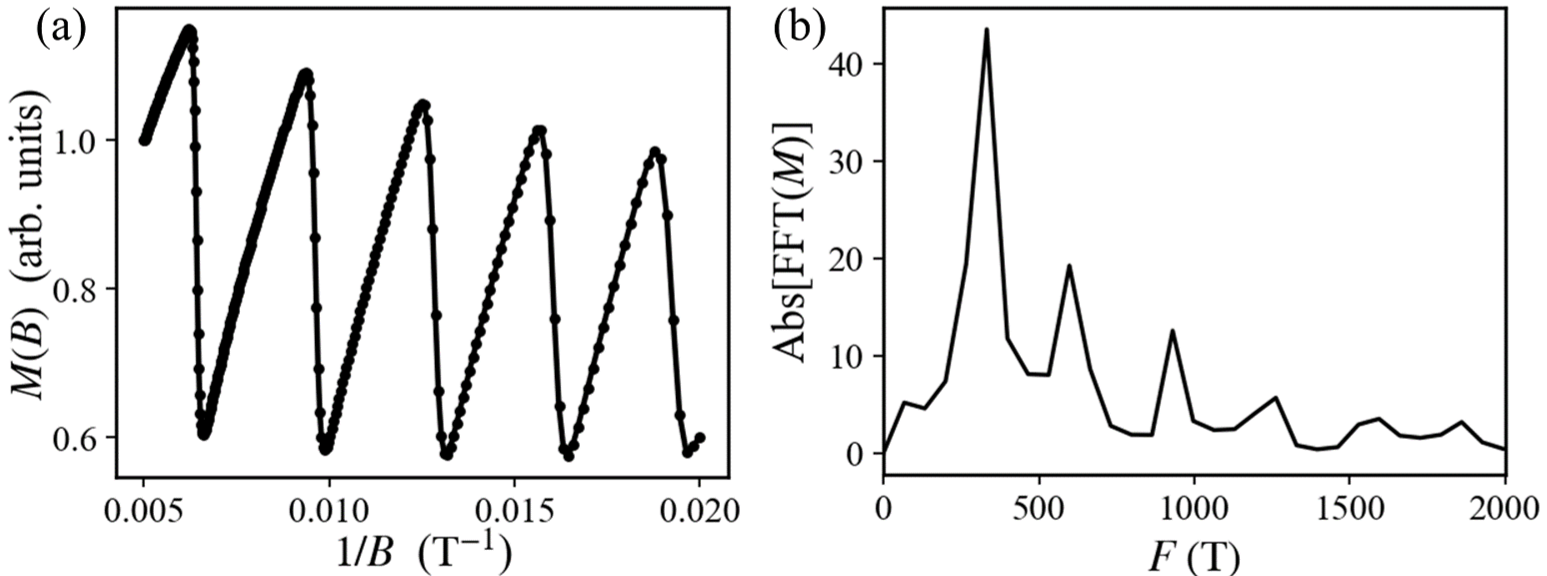}
    \caption{(a) Calculated magnetic polarization $M_z$ as a function of the inverse magnetic field $B^{-1}$. (b) Absolute value of Fourier transformed magnetic polarization}
    \label{fig:dhva}
\end{figure}

For completeness, we present here the full quantum-oscillation calculation based on the magnetic polarization, also called dHvA effect. We calculate the magnetic polarization with the following formula: $\Omega = -\tfrac{1}{\beta}\Tr\ln\left(-G^{-1}\right),\,M_z = -\frac{\partial\Omega}{\partial B_z}$, where $\beta=1/T$ is the inverse temperature and $\Omega$ is the grand potential. The Green's function used here is the same as Eq.~\ref{eq:GF} except that the interlayer hopping term is removed from the denominator. All the settings of parameters and magnetic field are the same as the section \textit{Quantum Oscillation}. The calculation results are presented in Fig.~\ref{fig:dhva}, in which the frequency $F=332\mathrm{T}$ and the period $\delta(1/B)=0.003\mathrm{T}^{-1}$ are the same as that in SdH results. Therefore, the dHvA calculation provides an independent and direct confirmation of the Fermi-pocket area.

\section{Spectrum function for Green's function with zeros}

\begin{figure}[thb]
    \centering
    \includegraphics[width=\linewidth]{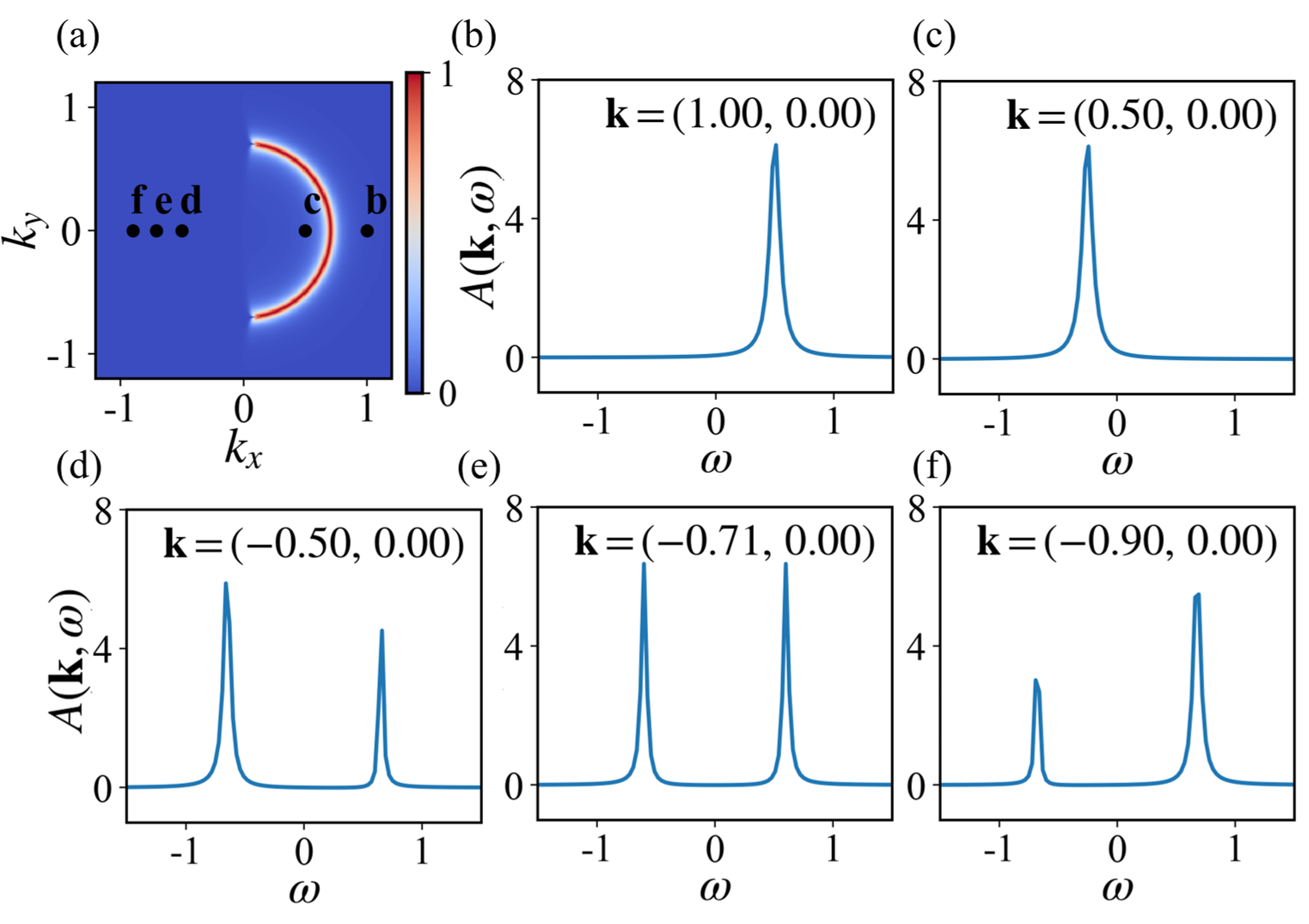}
    \caption{Energy distribution curve at different $\bfk$ points. 
    (a) Spectral function of the half suppressed case, with $\bfk$-points in panel (b)-(f) marked inside. EDC are shown on (b) $\bfk=(1,0)$,  (c) $\bfk=(0.5,0)$, (d) $\bfk=(-0.5,0)$, (e) $\bfk=(-k_F,0)$, and (f) $\bfk=(-1,0)$, with $k_F=1/\sqrt{2}$.}
    \label{fig:edc}
\end{figure}

In the main text, we analyze the role of zeros in the toy Green’s function $G_0$. Here, to further clarify their physical meaning, we examine the spectral function $A(k,\omega)$, which directly reveals the electronic character associated with Green’s function zeros. 

To illustrate this, we select five representative points along the $k_x$-axis in the half-circle case, as shown in Fig.~\ref{fig:edc}(a). Their corresponding $A(k,\omega)$ are displayed in Fig.~\ref{fig:edc}(b–f).
Points b and c, located on the right half of the pocket, cross a conventional Fermi surface. The peak in Fig.~\ref{fig:edc}(b) lies at $\omega>0$, indicating an unoccupied (hole-like) excitation. In Fig.~\ref{fig:edc}(c), the peak lies at $\omega<0$, identifying an occupied (electron-like) excitation. Thus, traditional Fermi points—corresponding to poles of the Green’s function—serve as sharp boundaries separating electron and hole character.

On the left half, however, the Fermi points have become zeros of the Green’s function. The corresponding $A(k,\omega)$ in Fig.~\ref{fig:edc}(d–f) exhibit both positive- and negative-energy peaks, signaling a mixed electron–hole character. At the zero itself, shown in Fig.~\ref{fig:edc}(e), the electron and hole contributions are equal, and the spectral function at $\omega=0$ vanishes—contrasting with the pronounced peak expected at an ordinary Fermi surface. This behavior highlights the fundamentally different electronic structure associated with Green’s function zeros.

\bibliography{refs}

\begin{thebibliography}{52}%
\makeatletter
\providecommand \@ifxundefined [1]{%
 \@ifx{#1\undefined}
}%
\providecommand \@ifnum [1]{%
 \ifnum #1\expandafter \@firstoftwo
 \else \expandafter \@secondoftwo
 \fi
}%
\providecommand \@ifx [1]{%
 \ifx #1\expandafter \@firstoftwo
 \else \expandafter \@secondoftwo
 \fi
}%
\providecommand \natexlab [1]{#1}%
\providecommand \enquote  [1]{``#1''}%
\providecommand \bibnamefont  [1]{#1}%
\providecommand \bibfnamefont [1]{#1}%
\providecommand \citenamefont [1]{#1}%
\providecommand \href@noop [0]{\@secondoftwo}%
\providecommand \href [0]{\begingroup \@sanitize@url \@href}%
\providecommand \@href[1]{\@@startlink{#1}\@@href}%
\providecommand \@@href[1]{\endgroup#1\@@endlink}%
\providecommand \@sanitize@url [0]{\catcode `\\12\catcode `\$12\catcode `\&12\catcode `\#12\catcode `\^12\catcode `\_12\catcode `\%12\relax}%
\providecommand \@@startlink[1]{}%
\providecommand \@@endlink[0]{}%
\providecommand \url  [0]{\begingroup\@sanitize@url \@url }%
\providecommand \@url [1]{\endgroup\@href {#1}{\urlprefix }}%
\providecommand \urlprefix  [0]{URL }%
\providecommand \Eprint [0]{\href }%
\providecommand \doibase [0]{http://dx.doi.org/}%
\providecommand \selectlanguage [0]{\@gobble}%
\providecommand \bibinfo  [0]{\@secondoftwo}%
\providecommand \bibfield  [0]{\@secondoftwo}%
\providecommand \translation [1]{[#1]}%
\providecommand \BibitemOpen [0]{}%
\providecommand \bibitemStop [0]{}%
\providecommand \bibitemNoStop [0]{.\EOS\space}%
\providecommand \EOS [0]{\spacefactor3000\relax}%
\providecommand \BibitemShut  [1]{\csname bibitem#1\endcsname}%
\let\auto@bib@innerbib\@empty
\bibitem [{\citenamefont {Anderson}\ \emph {et~al.}(2004)\citenamefont {Anderson}, \citenamefont {Lee}, \citenamefont {Randeria}, \citenamefont {Rice}, \citenamefont {Trivedi},\ and\ \citenamefont {Zhang}}]{Anderson_2004}%
  \BibitemOpen
  \bibfield  {author} {\bibinfo {author} {\bibfnamefont {P.~W.}\ \bibnamefont {Anderson}}, \bibinfo {author} {\bibfnamefont {P.~A.}\ \bibnamefont {Lee}}, \bibinfo {author} {\bibfnamefont {M.}~\bibnamefont {Randeria}}, \bibinfo {author} {\bibfnamefont {T.~M.}\ \bibnamefont {Rice}}, \bibinfo {author} {\bibfnamefont {N.}~\bibnamefont {Trivedi}}, \ and\ \bibinfo {author} {\bibfnamefont {F.~C.}\ \bibnamefont {Zhang}},\ }\href {\doibase 10.1088/0953-8984/16/24/R02} {\bibfield  {journal} {\bibinfo  {journal} {Journal of Physics: Condensed Matter}\ }\textbf {\bibinfo {volume} {16}},\ \bibinfo {pages} {R755} (\bibinfo {year} {2004})}\BibitemShut {NoStop}%
\bibitem [{\citenamefont {Keimer}\ \emph {et~al.}(2015)\citenamefont {Keimer}, \citenamefont {Kivelson}, \citenamefont {Norman}, \citenamefont {Uchida},\ and\ \citenamefont {Zaanen}}]{keimer_review}%
  \BibitemOpen
  \bibfield  {author} {\bibinfo {author} {\bibfnamefont {B.}~\bibnamefont {Keimer}}, \bibinfo {author} {\bibfnamefont {S.~A.}\ \bibnamefont {Kivelson}}, \bibinfo {author} {\bibfnamefont {M.~R.}\ \bibnamefont {Norman}}, \bibinfo {author} {\bibfnamefont {S.}~\bibnamefont {Uchida}}, \ and\ \bibinfo {author} {\bibfnamefont {J.}~\bibnamefont {Zaanen}},\ }\href {\doibase 10.1038/nature14165} {\bibfield  {journal} {\bibinfo  {journal} {Nature}\ }\textbf {\bibinfo {volume} {518}},\ \bibinfo {pages} {179} (\bibinfo {year} {2015})}\BibitemShut {NoStop}%
\bibitem [{\citenamefont {Timusk}\ and\ \citenamefont {Statt}(1999)}]{Timusk_1999}%
  \BibitemOpen
  \bibfield  {author} {\bibinfo {author} {\bibfnamefont {T.}~\bibnamefont {Timusk}}\ and\ \bibinfo {author} {\bibfnamefont {B.}~\bibnamefont {Statt}},\ }\href {\doibase 10.1088/0034-4885/62/1/002} {\bibfield  {journal} {\bibinfo  {journal} {Reports on Progress in Physics}\ }\textbf {\bibinfo {volume} {62}},\ \bibinfo {pages} {61} (\bibinfo {year} {1999})}\BibitemShut {NoStop}%
\bibitem [{\citenamefont {Rice}\ \emph {et~al.}(2011)\citenamefont {Rice}, \citenamefont {Yang},\ and\ \citenamefont {Zhang}}]{rice2011phenomenological}%
  \BibitemOpen
  \bibfield  {author} {\bibinfo {author} {\bibfnamefont {T.~M.}\ \bibnamefont {Rice}}, \bibinfo {author} {\bibfnamefont {K.-Y.}\ \bibnamefont {Yang}}, \ and\ \bibinfo {author} {\bibfnamefont {F.~C.}\ \bibnamefont {Zhang}},\ }\href {\doibase 10.1088/0034-4885/75/1/016502} {\bibfield  {journal} {\bibinfo  {journal} {Reports on Progress in Physics}\ }\textbf {\bibinfo {volume} {75}},\ \bibinfo {pages} {016502} (\bibinfo {year} {2011})}\BibitemShut {NoStop}%
\bibitem [{\citenamefont {Sebastian}\ \emph {et~al.}(2012)\citenamefont {Sebastian}, \citenamefont {Harrison},\ and\ \citenamefont {Lonzarich}}]{Sebastian_2012}%
  \BibitemOpen
  \bibfield  {author} {\bibinfo {author} {\bibfnamefont {S.~E.}\ \bibnamefont {Sebastian}}, \bibinfo {author} {\bibfnamefont {N.}~\bibnamefont {Harrison}}, \ and\ \bibinfo {author} {\bibfnamefont {G.~G.}\ \bibnamefont {Lonzarich}},\ }\href {\doibase 10.1088/0034-4885/75/10/102501} {\bibfield  {journal} {\bibinfo  {journal} {Reports on Progress in Physics}\ }\textbf {\bibinfo {volume} {75}},\ \bibinfo {pages} {102501} (\bibinfo {year} {2012})}\BibitemShut {NoStop}%
\bibitem [{\citenamefont {Warren}\ \emph {et~al.}(1989)\citenamefont {Warren}, \citenamefont {Walstedt}, \citenamefont {Brennert}, \citenamefont {Cava}, \citenamefont {Tycko}, \citenamefont {Bell},\ and\ \citenamefont {Dabbagh}}]{NMR_PhysRevLett.62.1193}%
  \BibitemOpen
  \bibfield  {author} {\bibinfo {author} {\bibfnamefont {W.~W.}\ \bibnamefont {Warren}}, \bibinfo {author} {\bibfnamefont {R.~E.}\ \bibnamefont {Walstedt}}, \bibinfo {author} {\bibfnamefont {G.~F.}\ \bibnamefont {Brennert}}, \bibinfo {author} {\bibfnamefont {R.~J.}\ \bibnamefont {Cava}}, \bibinfo {author} {\bibfnamefont {R.}~\bibnamefont {Tycko}}, \bibinfo {author} {\bibfnamefont {R.~F.}\ \bibnamefont {Bell}}, \ and\ \bibinfo {author} {\bibfnamefont {G.}~\bibnamefont {Dabbagh}},\ }\href {\doibase 10.1103/PhysRevLett.62.1193} {\bibfield  {journal} {\bibinfo  {journal} {Phys. Rev. Lett.}\ }\textbf {\bibinfo {volume} {62}},\ \bibinfo {pages} {1193} (\bibinfo {year} {1989})}\BibitemShut {NoStop}%
\bibitem [{\citenamefont {Alloul}\ \emph {et~al.}(1989)\citenamefont {Alloul}, \citenamefont {Ohno},\ and\ \citenamefont {Mendels}}]{NMR_PhysRevLett.63.1700}%
  \BibitemOpen
  \bibfield  {author} {\bibinfo {author} {\bibfnamefont {H.}~\bibnamefont {Alloul}}, \bibinfo {author} {\bibfnamefont {T.}~\bibnamefont {Ohno}}, \ and\ \bibinfo {author} {\bibfnamefont {P.}~\bibnamefont {Mendels}},\ }\href {\doibase 10.1103/PhysRevLett.63.1700} {\bibfield  {journal} {\bibinfo  {journal} {Phys. Rev. Lett.}\ }\textbf {\bibinfo {volume} {63}},\ \bibinfo {pages} {1700} (\bibinfo {year} {1989})}\BibitemShut {NoStop}%
\bibitem [{\citenamefont {Bucher}\ \emph {et~al.}(1993)\citenamefont {Bucher}, \citenamefont {Steiner}, \citenamefont {Karpinski}, \citenamefont {Kaldis},\ and\ \citenamefont {Wachter}}]{NMR_PhysRevLett.70.2012}%
  \BibitemOpen
  \bibfield  {author} {\bibinfo {author} {\bibfnamefont {B.}~\bibnamefont {Bucher}}, \bibinfo {author} {\bibfnamefont {P.}~\bibnamefont {Steiner}}, \bibinfo {author} {\bibfnamefont {J.}~\bibnamefont {Karpinski}}, \bibinfo {author} {\bibfnamefont {E.}~\bibnamefont {Kaldis}}, \ and\ \bibinfo {author} {\bibfnamefont {P.}~\bibnamefont {Wachter}},\ }\href {\doibase 10.1103/PhysRevLett.70.2012} {\bibfield  {journal} {\bibinfo  {journal} {Phys. Rev. Lett.}\ }\textbf {\bibinfo {volume} {70}},\ \bibinfo {pages} {2012} (\bibinfo {year} {1993})}\BibitemShut {NoStop}%
\bibitem [{\citenamefont {Puchkov}\ \emph {et~al.}(1996)\citenamefont {Puchkov}, \citenamefont {Basov},\ and\ \citenamefont {Timusk}}]{optical_1996}%
  \BibitemOpen
  \bibfield  {author} {\bibinfo {author} {\bibfnamefont {A.~V.}\ \bibnamefont {Puchkov}}, \bibinfo {author} {\bibfnamefont {D.~N.}\ \bibnamefont {Basov}}, \ and\ \bibinfo {author} {\bibfnamefont {T.}~\bibnamefont {Timusk}},\ }\href {\doibase 10.1088/0953-8984/8/48/023} {\bibfield  {journal} {\bibinfo  {journal} {Journal of Physics: Condensed Matter}\ }\textbf {\bibinfo {volume} {8}},\ \bibinfo {pages} {10049} (\bibinfo {year} {1996})}\BibitemShut {NoStop}%
\bibitem [{\citenamefont {Homes}\ \emph {et~al.}(1993)\citenamefont {Homes}, \citenamefont {Timusk}, \citenamefont {Liang}, \citenamefont {Bonn},\ and\ \citenamefont {Hardy}}]{optical_PhysRevLett.71.1645}%
  \BibitemOpen
  \bibfield  {author} {\bibinfo {author} {\bibfnamefont {C.~C.}\ \bibnamefont {Homes}}, \bibinfo {author} {\bibfnamefont {T.}~\bibnamefont {Timusk}}, \bibinfo {author} {\bibfnamefont {R.}~\bibnamefont {Liang}}, \bibinfo {author} {\bibfnamefont {D.~A.}\ \bibnamefont {Bonn}}, \ and\ \bibinfo {author} {\bibfnamefont {W.~N.}\ \bibnamefont {Hardy}},\ }\href {\doibase 10.1103/PhysRevLett.71.1645} {\bibfield  {journal} {\bibinfo  {journal} {Phys. Rev. Lett.}\ }\textbf {\bibinfo {volume} {71}},\ \bibinfo {pages} {1645} (\bibinfo {year} {1993})}\BibitemShut {NoStop}%
\bibitem [{\citenamefont {Ito}\ \emph {et~al.}(1993)\citenamefont {Ito}, \citenamefont {Takenaka},\ and\ \citenamefont {Uchida}}]{Transport_PhysRevLett.70.3995}%
  \BibitemOpen
  \bibfield  {author} {\bibinfo {author} {\bibfnamefont {T.}~\bibnamefont {Ito}}, \bibinfo {author} {\bibfnamefont {K.}~\bibnamefont {Takenaka}}, \ and\ \bibinfo {author} {\bibfnamefont {S.}~\bibnamefont {Uchida}},\ }\href {\doibase 10.1103/PhysRevLett.70.3995} {\bibfield  {journal} {\bibinfo  {journal} {Phys. Rev. Lett.}\ }\textbf {\bibinfo {volume} {70}},\ \bibinfo {pages} {3995} (\bibinfo {year} {1993})}\BibitemShut {NoStop}%
\bibitem [{\citenamefont {Marshall}\ \emph {et~al.}(1996)\citenamefont {Marshall}, \citenamefont {Dessau}, \citenamefont {Loeser}, \citenamefont {Park}, \citenamefont {Matsuura}, \citenamefont {Eckstein}, \citenamefont {Bozovic}, \citenamefont {Fournier}, \citenamefont {Kapitulnik}, \citenamefont {Spicer},\ and\ \citenamefont {Shen}}]{Shen_PhysRevLett.76.4841}%
  \BibitemOpen
  \bibfield  {author} {\bibinfo {author} {\bibfnamefont {D.~S.}\ \bibnamefont {Marshall}}, \bibinfo {author} {\bibfnamefont {D.~S.}\ \bibnamefont {Dessau}}, \bibinfo {author} {\bibfnamefont {A.~G.}\ \bibnamefont {Loeser}}, \bibinfo {author} {\bibfnamefont {C.-H.}\ \bibnamefont {Park}}, \bibinfo {author} {\bibfnamefont {A.~Y.}\ \bibnamefont {Matsuura}}, \bibinfo {author} {\bibfnamefont {J.~N.}\ \bibnamefont {Eckstein}}, \bibinfo {author} {\bibfnamefont {I.}~\bibnamefont {Bozovic}}, \bibinfo {author} {\bibfnamefont {P.}~\bibnamefont {Fournier}}, \bibinfo {author} {\bibfnamefont {A.}~\bibnamefont {Kapitulnik}}, \bibinfo {author} {\bibfnamefont {W.~E.}\ \bibnamefont {Spicer}}, \ and\ \bibinfo {author} {\bibfnamefont {Z.-X.}\ \bibnamefont {Shen}},\ }\href {\doibase 10.1103/PhysRevLett.76.4841} {\bibfield  {journal} {\bibinfo  {journal} {Phys. Rev. Lett.}\ }\textbf {\bibinfo {volume} {76}},\ \bibinfo {pages} {4841} (\bibinfo {year} {1996})}\BibitemShut {NoStop}%
\bibitem [{\citenamefont {Ding}\ \emph {et~al.}(1996)\citenamefont {Ding}, \citenamefont {Yokoya}, \citenamefont {Campuzano}, \citenamefont {Takahashi}, \citenamefont {Randeria}, \citenamefont {Norman}, \citenamefont {Mochiku}, \citenamefont {Kadowaki},\ and\ \citenamefont {Giapintzakis}}]{ding_1996}%
  \BibitemOpen
  \bibfield  {author} {\bibinfo {author} {\bibfnamefont {H.}~\bibnamefont {Ding}}, \bibinfo {author} {\bibfnamefont {T.}~\bibnamefont {Yokoya}}, \bibinfo {author} {\bibfnamefont {J.~C.}\ \bibnamefont {Campuzano}}, \bibinfo {author} {\bibfnamefont {T.}~\bibnamefont {Takahashi}}, \bibinfo {author} {\bibfnamefont {M.}~\bibnamefont {Randeria}}, \bibinfo {author} {\bibfnamefont {M.~R.}\ \bibnamefont {Norman}}, \bibinfo {author} {\bibfnamefont {T.}~\bibnamefont {Mochiku}}, \bibinfo {author} {\bibfnamefont {K.}~\bibnamefont {Kadowaki}}, \ and\ \bibinfo {author} {\bibfnamefont {J.}~\bibnamefont {Giapintzakis}},\ }\href {\doibase 10.1038/382051a0} {\bibfield  {journal} {\bibinfo  {journal} {Nature}\ }\textbf {\bibinfo {volume} {382}},\ \bibinfo {pages} {51} (\bibinfo {year} {1996})}\BibitemShut {NoStop}%
\bibitem [{\citenamefont {Norman}\ \emph {et~al.}(1998)\citenamefont {Norman}, \citenamefont {Ding}, \citenamefont {Randeria}, \citenamefont {Campuzano}, \citenamefont {Yokoya}, \citenamefont {Takeuchi}, \citenamefont {Takahashi}, \citenamefont {Mochiku}, \citenamefont {Kadowaki}, \citenamefont {Guptasarma},\ and\ \citenamefont {Hinks}}]{norman1998destruction}%
  \BibitemOpen
  \bibfield  {author} {\bibinfo {author} {\bibfnamefont {M.~R.}\ \bibnamefont {Norman}}, \bibinfo {author} {\bibfnamefont {H.}~\bibnamefont {Ding}}, \bibinfo {author} {\bibfnamefont {M.}~\bibnamefont {Randeria}}, \bibinfo {author} {\bibfnamefont {J.~C.}\ \bibnamefont {Campuzano}}, \bibinfo {author} {\bibfnamefont {T.}~\bibnamefont {Yokoya}}, \bibinfo {author} {\bibfnamefont {T.}~\bibnamefont {Takeuchi}}, \bibinfo {author} {\bibfnamefont {T.}~\bibnamefont {Takahashi}}, \bibinfo {author} {\bibfnamefont {T.}~\bibnamefont {Mochiku}}, \bibinfo {author} {\bibfnamefont {K.}~\bibnamefont {Kadowaki}}, \bibinfo {author} {\bibfnamefont {P.}~\bibnamefont {Guptasarma}}, \ and\ \bibinfo {author} {\bibfnamefont {D.~G.}\ \bibnamefont {Hinks}},\ }\href {\doibase 10.1038/32366} {\bibfield  {journal} {\bibinfo  {journal} {Nature}\ }\textbf {\bibinfo {volume} {392}},\ \bibinfo {pages} {157} (\bibinfo {year} {1998})}\BibitemShut {NoStop}%
\bibitem [{\citenamefont {Shen}\ \emph {et~al.}(2005)\citenamefont {Shen}, \citenamefont {Ronning}, \citenamefont {Lu}, \citenamefont {Baumberger}, \citenamefont {Ingle}, \citenamefont {Lee}, \citenamefont {Meevasana}, \citenamefont {Kohsaka}, \citenamefont {Azuma}, \citenamefont {Takano}, \citenamefont {Takagi},\ and\ \citenamefont {Shen}}]{kyle_shen}%
  \BibitemOpen
  \bibfield  {author} {\bibinfo {author} {\bibfnamefont {K.~M.}\ \bibnamefont {Shen}}, \bibinfo {author} {\bibfnamefont {F.}~\bibnamefont {Ronning}}, \bibinfo {author} {\bibfnamefont {D.~H.}\ \bibnamefont {Lu}}, \bibinfo {author} {\bibfnamefont {F.}~\bibnamefont {Baumberger}}, \bibinfo {author} {\bibfnamefont {N.~J.~C.}\ \bibnamefont {Ingle}}, \bibinfo {author} {\bibfnamefont {W.~S.}\ \bibnamefont {Lee}}, \bibinfo {author} {\bibfnamefont {W.}~\bibnamefont {Meevasana}}, \bibinfo {author} {\bibfnamefont {Y.}~\bibnamefont {Kohsaka}}, \bibinfo {author} {\bibfnamefont {M.}~\bibnamefont {Azuma}}, \bibinfo {author} {\bibfnamefont {M.}~\bibnamefont {Takano}}, \bibinfo {author} {\bibfnamefont {H.}~\bibnamefont {Takagi}}, \ and\ \bibinfo {author} {\bibfnamefont {Z.-X.}\ \bibnamefont {Shen}},\ }\href {\doibase 10.1126/science.1103627} {\bibfield  {journal} {\bibinfo  {journal} {Science}\ }\textbf {\bibinfo {volume} {307}},\ \bibinfo {pages} {901} (\bibinfo {year} {2005})}\BibitemShut {NoStop}%
\bibitem [{\citenamefont {Yang}\ \emph {et~al.}(2011)\citenamefont {Yang}, \citenamefont {Rameau}, \citenamefont {Pan}, \citenamefont {Gu}, \citenamefont {Johnson}, \citenamefont {Claus}, \citenamefont {Hinks},\ and\ \citenamefont {Kidd}}]{HBYang_PhysRevLett.107.047003}%
  \BibitemOpen
  \bibfield  {author} {\bibinfo {author} {\bibfnamefont {H.-B.}\ \bibnamefont {Yang}}, \bibinfo {author} {\bibfnamefont {J.~D.}\ \bibnamefont {Rameau}}, \bibinfo {author} {\bibfnamefont {Z.-H.}\ \bibnamefont {Pan}}, \bibinfo {author} {\bibfnamefont {G.~D.}\ \bibnamefont {Gu}}, \bibinfo {author} {\bibfnamefont {P.~D.}\ \bibnamefont {Johnson}}, \bibinfo {author} {\bibfnamefont {H.}~\bibnamefont {Claus}}, \bibinfo {author} {\bibfnamefont {D.~G.}\ \bibnamefont {Hinks}}, \ and\ \bibinfo {author} {\bibfnamefont {T.~E.}\ \bibnamefont {Kidd}},\ }\href {\doibase 10.1103/PhysRevLett.107.047003} {\bibfield  {journal} {\bibinfo  {journal} {Phys. Rev. Lett.}\ }\textbf {\bibinfo {volume} {107}},\ \bibinfo {pages} {047003} (\bibinfo {year} {2011})}\BibitemShut {NoStop}%
\bibitem [{\citenamefont {Hashimoto}\ \emph {et~al.}(2009)\citenamefont {Hashimoto}, \citenamefont {Yoshida}, \citenamefont {Tanaka}, \citenamefont {Fujimori}, \citenamefont {Okusawa}, \citenamefont {Wakimoto}, \citenamefont {Yamada}, \citenamefont {Kakeshita}, \citenamefont {Eisaki},\ and\ \citenamefont {Uchida}}]{Uchida_PhysRevB.79.140502}%
  \BibitemOpen
  \bibfield  {author} {\bibinfo {author} {\bibfnamefont {M.}~\bibnamefont {Hashimoto}}, \bibinfo {author} {\bibfnamefont {T.}~\bibnamefont {Yoshida}}, \bibinfo {author} {\bibfnamefont {K.}~\bibnamefont {Tanaka}}, \bibinfo {author} {\bibfnamefont {A.}~\bibnamefont {Fujimori}}, \bibinfo {author} {\bibfnamefont {M.}~\bibnamefont {Okusawa}}, \bibinfo {author} {\bibfnamefont {S.}~\bibnamefont {Wakimoto}}, \bibinfo {author} {\bibfnamefont {K.}~\bibnamefont {Yamada}}, \bibinfo {author} {\bibfnamefont {T.}~\bibnamefont {Kakeshita}}, \bibinfo {author} {\bibfnamefont {H.}~\bibnamefont {Eisaki}}, \ and\ \bibinfo {author} {\bibfnamefont {S.}~\bibnamefont {Uchida}},\ }\href {\doibase 10.1103/PhysRevB.79.140502} {\bibfield  {journal} {\bibinfo  {journal} {Phys. Rev. B}\ }\textbf {\bibinfo {volume} {79}},\ \bibinfo {pages} {140502} (\bibinfo {year} {2009})}\BibitemShut {NoStop}%
\bibitem [{\citenamefont {Yang}\ \emph {et~al.}(2008{\natexlab{a}})\citenamefont {Yang}, \citenamefont {Rameau}, \citenamefont {Johnson}, \citenamefont {Valla}, \citenamefont {Tsvelik},\ and\ \citenamefont {Gu}}]{HBYang_nature}%
  \BibitemOpen
  \bibfield  {author} {\bibinfo {author} {\bibfnamefont {H.~B.}\ \bibnamefont {Yang}}, \bibinfo {author} {\bibfnamefont {J.~D.}\ \bibnamefont {Rameau}}, \bibinfo {author} {\bibfnamefont {P.~D.}\ \bibnamefont {Johnson}}, \bibinfo {author} {\bibfnamefont {T.}~\bibnamefont {Valla}}, \bibinfo {author} {\bibfnamefont {A.}~\bibnamefont {Tsvelik}}, \ and\ \bibinfo {author} {\bibfnamefont {G.~D.}\ \bibnamefont {Gu}},\ }\href {\doibase 10.1038/nature07400} {\bibfield  {journal} {\bibinfo  {journal} {Nature}\ }\textbf {\bibinfo {volume} {456}},\ \bibinfo {pages} {77} (\bibinfo {year} {2008}{\natexlab{a}})}\BibitemShut {NoStop}%
\bibitem [{\citenamefont {Yang}\ \emph {et~al.}(2006)\citenamefont {Yang}, \citenamefont {Rice},\ and\ \citenamefont {Zhang}}]{YRZ}%
  \BibitemOpen
  \bibfield  {author} {\bibinfo {author} {\bibfnamefont {K.-Y.}\ \bibnamefont {Yang}}, \bibinfo {author} {\bibfnamefont {T.~M.}\ \bibnamefont {Rice}}, \ and\ \bibinfo {author} {\bibfnamefont {F.-C.}\ \bibnamefont {Zhang}},\ }\href {\doibase 10.1103/PhysRevB.73.174501} {\bibfield  {journal} {\bibinfo  {journal} {Phys. Rev. B}\ }\textbf {\bibinfo {volume} {73}},\ \bibinfo {pages} {174501} (\bibinfo {year} {2006})}\BibitemShut {NoStop}%
\bibitem [{\citenamefont {Chen}\ \emph {et~al.}(2008)\citenamefont {Chen}, \citenamefont {Yang}, \citenamefont {Rice},\ and\ \citenamefont {Zhang}}]{Chen_2008}%
  \BibitemOpen
  \bibfield  {author} {\bibinfo {author} {\bibfnamefont {W.-Q.}\ \bibnamefont {Chen}}, \bibinfo {author} {\bibfnamefont {K.-Y.}\ \bibnamefont {Yang}}, \bibinfo {author} {\bibfnamefont {T.~M.}\ \bibnamefont {Rice}}, \ and\ \bibinfo {author} {\bibfnamefont {F.~C.}\ \bibnamefont {Zhang}},\ }\href {\doibase 10.1209/0295-5075/82/17004} {\bibfield  {journal} {\bibinfo  {journal} {Europhysics Letters}\ }\textbf {\bibinfo {volume} {82}},\ \bibinfo {pages} {17004} (\bibinfo {year} {2008})}\BibitemShut {NoStop}%
\bibitem [{\citenamefont {Yang}\ \emph {et~al.}(2008{\natexlab{b}})\citenamefont {Yang}, \citenamefont {Yang}, \citenamefont {Johnson}, \citenamefont {Rice},\ and\ \citenamefont {Zhang}}]{Yang_2009}%
  \BibitemOpen
  \bibfield  {author} {\bibinfo {author} {\bibfnamefont {K.-Y.}\ \bibnamefont {Yang}}, \bibinfo {author} {\bibfnamefont {H.-B.}\ \bibnamefont {Yang}}, \bibinfo {author} {\bibfnamefont {P.~D.}\ \bibnamefont {Johnson}}, \bibinfo {author} {\bibfnamefont {T.~M.}\ \bibnamefont {Rice}}, \ and\ \bibinfo {author} {\bibfnamefont {F.-C.}\ \bibnamefont {Zhang}},\ }\href {\doibase 10.1209/0295-5075/86/37002} {\bibfield  {journal} {\bibinfo  {journal} {Europhysics Letters}\ }\textbf {\bibinfo {volume} {86}},\ \bibinfo {pages} {37002} (\bibinfo {year} {2008}{\natexlab{b}})}\BibitemShut {NoStop}%
\bibitem [{\citenamefont {Rice}\ \emph {et~al.}(2017)\citenamefont {Rice}, \citenamefont {Robinson},\ and\ \citenamefont {Tsvelik}}]{Rice2017PhysRevB.96.220502}%
  \BibitemOpen
  \bibfield  {author} {\bibinfo {author} {\bibfnamefont {T.~M.}\ \bibnamefont {Rice}}, \bibinfo {author} {\bibfnamefont {N.~J.}\ \bibnamefont {Robinson}}, \ and\ \bibinfo {author} {\bibfnamefont {A.~M.}\ \bibnamefont {Tsvelik}},\ }\href {\doibase 10.1103/PhysRevB.96.220502} {\bibfield  {journal} {\bibinfo  {journal} {Phys. Rev. B}\ }\textbf {\bibinfo {volume} {96}},\ \bibinfo {pages} {220502} (\bibinfo {year} {2017})}\BibitemShut {NoStop}%
\bibitem [{\citenamefont {Zhang}\ and\ \citenamefont {Mei}(2016)}]{Zhanglong}%
  \BibitemOpen
  \bibfield  {author} {\bibinfo {author} {\bibfnamefont {L.}~\bibnamefont {Zhang}}\ and\ \bibinfo {author} {\bibfnamefont {J.-W.}\ \bibnamefont {Mei}},\ }\href {\doibase 10.1209/0295-5075/114/47008} {\bibfield  {journal} {\bibinfo  {journal} {Europhysics Letters}\ }\textbf {\bibinfo {volume} {114}},\ \bibinfo {pages} {47008} (\bibinfo {year} {2016})}\BibitemShut {NoStop}%
\bibitem [{\citenamefont {Anderson}(1987)}]{anderson1987resonating}%
  \BibitemOpen
  \bibfield  {author} {\bibinfo {author} {\bibfnamefont {P.~W.}\ \bibnamefont {Anderson}},\ }\href {\doibase 10.1126/science.235.4793.1196} {\bibfield  {journal} {\bibinfo  {journal} {Science}\ }\textbf {\bibinfo {volume} {235}},\ \bibinfo {pages} {1196} (\bibinfo {year} {1987})}\BibitemShut {NoStop}%
\bibitem [{\citenamefont {Zhang}\ and\ \citenamefont {Rice}(1988)}]{zhang1988effective}%
  \BibitemOpen
  \bibfield  {author} {\bibinfo {author} {\bibfnamefont {F.~C.}\ \bibnamefont {Zhang}}\ and\ \bibinfo {author} {\bibfnamefont {T.~M.}\ \bibnamefont {Rice}},\ }\href {\doibase 10.1103/PhysRevB.37.3759} {\bibfield  {journal} {\bibinfo  {journal} {Phys. Rev. B}\ }\textbf {\bibinfo {volume} {37}},\ \bibinfo {pages} {3759} (\bibinfo {year} {1988})}\BibitemShut {NoStop}%
\bibitem [{\citenamefont {Zhang}\ \emph {et~al.}(1988)\citenamefont {Zhang}, \citenamefont {Gros}, \citenamefont {Rice},\ and\ \citenamefont {Shiba}}]{zhang1988renormalised}%
  \BibitemOpen
  \bibfield  {author} {\bibinfo {author} {\bibfnamefont {F.~C.}\ \bibnamefont {Zhang}}, \bibinfo {author} {\bibfnamefont {C.}~\bibnamefont {Gros}}, \bibinfo {author} {\bibfnamefont {T.~M.}\ \bibnamefont {Rice}}, \ and\ \bibinfo {author} {\bibfnamefont {H.}~\bibnamefont {Shiba}},\ }\href {\doibase 10.1088/0953-2048/1/1/009} {\bibfield  {journal} {\bibinfo  {journal} {Superconductor Science and Technology}\ }\textbf {\bibinfo {volume} {1}},\ \bibinfo {pages} {36} (\bibinfo {year} {1988})}\BibitemShut {NoStop}%
\bibitem [{\citenamefont {Kohsaka}\ \emph {et~al.}(2008)\citenamefont {Kohsaka}, \citenamefont {Taylor}, \citenamefont {Wahl}, \citenamefont {Schmidt}, \citenamefont {Lee}, \citenamefont {Fujita}, \citenamefont {Alldredge}, \citenamefont {McElroy}, \citenamefont {Lee}, \citenamefont {Eisaki}, \citenamefont {Uchida}, \citenamefont {Lee},\ and\ \citenamefont {Davis}}]{davis_stm}%
  \BibitemOpen
  \bibfield  {author} {\bibinfo {author} {\bibfnamefont {Y.}~\bibnamefont {Kohsaka}}, \bibinfo {author} {\bibfnamefont {C.}~\bibnamefont {Taylor}}, \bibinfo {author} {\bibfnamefont {P.}~\bibnamefont {Wahl}}, \bibinfo {author} {\bibfnamefont {A.}~\bibnamefont {Schmidt}}, \bibinfo {author} {\bibfnamefont {J.}~\bibnamefont {Lee}}, \bibinfo {author} {\bibfnamefont {K.}~\bibnamefont {Fujita}}, \bibinfo {author} {\bibfnamefont {J.~W.}\ \bibnamefont {Alldredge}}, \bibinfo {author} {\bibfnamefont {K.}~\bibnamefont {McElroy}}, \bibinfo {author} {\bibfnamefont {J.}~\bibnamefont {Lee}}, \bibinfo {author} {\bibfnamefont {H.}~\bibnamefont {Eisaki}}, \bibinfo {author} {\bibfnamefont {S.}~\bibnamefont {Uchida}}, \bibinfo {author} {\bibfnamefont {D.~H.}\ \bibnamefont {Lee}}, \ and\ \bibinfo {author} {\bibfnamefont {J.~C.}\ \bibnamefont {Davis}},\ }\href {\doibase 10.1038/nature07243} {\bibfield  {journal} {\bibinfo  {journal} {Nature}\ }\textbf {\bibinfo {volume} {454}},\ \bibinfo {pages} {1072} (\bibinfo {year}
  {2008})}\BibitemShut {NoStop}%
\bibitem [{\citenamefont {Kanigel}\ \emph {et~al.}(2006)\citenamefont {Kanigel}, \citenamefont {Norman}, \citenamefont {Randeria}, \citenamefont {Chatterjee}, \citenamefont {Souma}, \citenamefont {Kaminski}, \citenamefont {Fretwell}, \citenamefont {Rosenkranz}, \citenamefont {Shi}, \citenamefont {Sato} \emph {et~al.}}]{kanigel2006evolution}%
  \BibitemOpen
  \bibfield  {author} {\bibinfo {author} {\bibfnamefont {A.}~\bibnamefont {Kanigel}}, \bibinfo {author} {\bibfnamefont {M.}~\bibnamefont {Norman}}, \bibinfo {author} {\bibfnamefont {M.}~\bibnamefont {Randeria}}, \bibinfo {author} {\bibfnamefont {U.}~\bibnamefont {Chatterjee}}, \bibinfo {author} {\bibfnamefont {S.}~\bibnamefont {Souma}}, \bibinfo {author} {\bibfnamefont {A.}~\bibnamefont {Kaminski}}, \bibinfo {author} {\bibfnamefont {H.}~\bibnamefont {Fretwell}}, \bibinfo {author} {\bibfnamefont {S.}~\bibnamefont {Rosenkranz}}, \bibinfo {author} {\bibfnamefont {M.}~\bibnamefont {Shi}}, \bibinfo {author} {\bibfnamefont {T.}~\bibnamefont {Sato}},  \emph {et~al.},\ }\href {https://doi.org/10.1038/nphys334} {\bibfield  {journal} {\bibinfo  {journal} {Nature Physics}\ }\textbf {\bibinfo {volume} {2}},\ \bibinfo {pages} {447} (\bibinfo {year} {2006})}\BibitemShut {NoStop}%
\bibitem [{\citenamefont {Stanescu}\ \emph {et~al.}(2007)\citenamefont {Stanescu}, \citenamefont {Phillips},\ and\ \citenamefont {Choy}}]{Luttingersurface}%
  \BibitemOpen
  \bibfield  {author} {\bibinfo {author} {\bibfnamefont {T.~D.}\ \bibnamefont {Stanescu}}, \bibinfo {author} {\bibfnamefont {P.}~\bibnamefont {Phillips}}, \ and\ \bibinfo {author} {\bibfnamefont {T.-P.}\ \bibnamefont {Choy}},\ }\href {\doibase 10.1103/PhysRevB.75.104503} {\bibfield  {journal} {\bibinfo  {journal} {Phys. Rev. B}\ }\textbf {\bibinfo {volume} {75}},\ \bibinfo {pages} {104503} (\bibinfo {year} {2007})}\BibitemShut {NoStop}%
\bibitem [{\citenamefont {Chan}\ \emph {et~al.}(2025)\citenamefont {Chan}, \citenamefont {Schreiber}, \citenamefont {Ayala-Valenzuela}, \citenamefont {Bauer}, \citenamefont {Shekhter},\ and\ \citenamefont {Harrison}}]{chan2025observation}%
  \BibitemOpen
  \bibfield  {author} {\bibinfo {author} {\bibfnamefont {M.~K.}\ \bibnamefont {Chan}}, \bibinfo {author} {\bibfnamefont {K.~A.}\ \bibnamefont {Schreiber}}, \bibinfo {author} {\bibfnamefont {O.~E.}\ \bibnamefont {Ayala-Valenzuela}}, \bibinfo {author} {\bibfnamefont {E.~D.}\ \bibnamefont {Bauer}}, \bibinfo {author} {\bibfnamefont {A.}~\bibnamefont {Shekhter}}, \ and\ \bibinfo {author} {\bibfnamefont {N.}~\bibnamefont {Harrison}},\ }\href {https://doi.org/10.1038/s41567-025-03032-2} {\bibfield  {journal} {\bibinfo  {journal} {Nature Physics}\ ,\ \bibinfo {pages} {1}} (\bibinfo {year} {2025})}\BibitemShut {NoStop}%
\bibitem [{\citenamefont {Yamaji}(1989)}]{yamaji1989angle}%
  \BibitemOpen
  \bibfield  {author} {\bibinfo {author} {\bibfnamefont {K.}~\bibnamefont {Yamaji}},\ }\href {https://journals.jps.jp/doi/10.1143/JPSJ.58.1520} {\bibfield  {journal} {\bibinfo  {journal} {Journal of the Physical Society of Japan}\ }\textbf {\bibinfo {volume} {58}},\ \bibinfo {pages} {1520} (\bibinfo {year} {1989})}\BibitemShut {NoStop}%
\bibitem [{\citenamefont {Yagi}\ \emph {et~al.}(1990)\citenamefont {Yagi}, \citenamefont {Iye}, \citenamefont {Osada},\ and\ \citenamefont {Kagoshima}}]{yagi1990semiclassical}%
  \BibitemOpen
  \bibfield  {author} {\bibinfo {author} {\bibfnamefont {R.}~\bibnamefont {Yagi}}, \bibinfo {author} {\bibfnamefont {Y.}~\bibnamefont {Iye}}, \bibinfo {author} {\bibfnamefont {T.}~\bibnamefont {Osada}}, \ and\ \bibinfo {author} {\bibfnamefont {S.}~\bibnamefont {Kagoshima}},\ }\href {https://journals.jps.jp/doi/10.1143/JPSJ.59.3069} {\bibfield  {journal} {\bibinfo  {journal} {Journal of the Physical Society of Japan}\ }\textbf {\bibinfo {volume} {59}},\ \bibinfo {pages} {3069} (\bibinfo {year} {1990})}\BibitemShut {NoStop}%
\bibitem [{\citenamefont {Kurihara}(1992)}]{kurihara1992microscopic}%
  \BibitemOpen
  \bibfield  {author} {\bibinfo {author} {\bibfnamefont {Y.}~\bibnamefont {Kurihara}},\ }\href {\doibase 10.1143/JPSJ.61.975} {\bibfield  {journal} {\bibinfo  {journal} {Journal of the Physical Society of Japan}\ }\textbf {\bibinfo {volume} {61}},\ \bibinfo {pages} {975} (\bibinfo {year} {1992})}\BibitemShut {NoStop}%
\bibitem [{\citenamefont {Fang}\ \emph {et~al.}(2022)\citenamefont {Fang}, \citenamefont {Grissonnanche}, \citenamefont {Legros}, \citenamefont {Verret}, \citenamefont {Lalibert{\'e}}, \citenamefont {Collignon}, \citenamefont {Ataei}, \citenamefont {Dion}, \citenamefont {Zhou}, \citenamefont {Graf} \emph {et~al.}}]{fang2022fermi}%
  \BibitemOpen
  \bibfield  {author} {\bibinfo {author} {\bibfnamefont {Y.}~\bibnamefont {Fang}}, \bibinfo {author} {\bibfnamefont {G.}~\bibnamefont {Grissonnanche}}, \bibinfo {author} {\bibfnamefont {A.}~\bibnamefont {Legros}}, \bibinfo {author} {\bibfnamefont {S.}~\bibnamefont {Verret}}, \bibinfo {author} {\bibfnamefont {F.}~\bibnamefont {Lalibert{\'e}}}, \bibinfo {author} {\bibfnamefont {C.}~\bibnamefont {Collignon}}, \bibinfo {author} {\bibfnamefont {A.}~\bibnamefont {Ataei}}, \bibinfo {author} {\bibfnamefont {M.}~\bibnamefont {Dion}}, \bibinfo {author} {\bibfnamefont {J.}~\bibnamefont {Zhou}}, \bibinfo {author} {\bibfnamefont {D.}~\bibnamefont {Graf}},  \emph {et~al.},\ }\href {https://doi.org/10.1038/s41567-022-01514-1} {\bibfield  {journal} {\bibinfo  {journal} {Nature Physics}\ }\textbf {\bibinfo {volume} {18}},\ \bibinfo {pages} {558} (\bibinfo {year} {2022})}\BibitemShut {NoStop}%
\bibitem [{\citenamefont {Zhao}\ \emph {et~al.}(2025)\citenamefont {Zhao}, \citenamefont {Chatterjee}, \citenamefont {Sachdev},\ and\ \citenamefont {Zhang}}]{zhao2025yamaji}%
  \BibitemOpen
  \bibfield  {author} {\bibinfo {author} {\bibfnamefont {J.-Y.}\ \bibnamefont {Zhao}}, \bibinfo {author} {\bibfnamefont {S.}~\bibnamefont {Chatterjee}}, \bibinfo {author} {\bibfnamefont {S.}~\bibnamefont {Sachdev}}, \ and\ \bibinfo {author} {\bibfnamefont {Y.-H.}\ \bibnamefont {Zhang}},\ }\href {https://doi.org/10.48550/arXiv.2510.13943} {\enquote {\bibinfo {title} {Yamaji effect in models of underdoped cuprates},}\ } (\bibinfo {year} {2025}),\ \Eprint {http://arxiv.org/abs/2510.13943} {arXiv:2510.13943 [cond-mat.str-el]} \BibitemShut {NoStop}%
\bibitem [{\citenamefont {Senthil}\ \emph {et~al.}(2003)\citenamefont {Senthil}, \citenamefont {Sachdev},\ and\ \citenamefont {Vojta}}]{sachdev2003}%
  \BibitemOpen
  \bibfield  {author} {\bibinfo {author} {\bibfnamefont {T.}~\bibnamefont {Senthil}}, \bibinfo {author} {\bibfnamefont {S.}~\bibnamefont {Sachdev}}, \ and\ \bibinfo {author} {\bibfnamefont {M.}~\bibnamefont {Vojta}},\ }\href {\doibase 10.1103/PhysRevLett.90.216403} {\bibfield  {journal} {\bibinfo  {journal} {Phys. Rev. Lett.}\ }\textbf {\bibinfo {volume} {90}},\ \bibinfo {pages} {216403} (\bibinfo {year} {2003})}\BibitemShut {NoStop}%
\bibitem [{\citenamefont {Qi}\ and\ \citenamefont {Sachdev}(2010)}]{sachdev2010}%
  \BibitemOpen
  \bibfield  {author} {\bibinfo {author} {\bibfnamefont {Y.}~\bibnamefont {Qi}}\ and\ \bibinfo {author} {\bibfnamefont {S.}~\bibnamefont {Sachdev}},\ }\href {\doibase 10.1103/PhysRevB.81.115129} {\bibfield  {journal} {\bibinfo  {journal} {Phys. Rev. B}\ }\textbf {\bibinfo {volume} {81}},\ \bibinfo {pages} {115129} (\bibinfo {year} {2010})}\BibitemShut {NoStop}%
\bibitem [{\citenamefont {Chatterjee}\ and\ \citenamefont {Sachdev}(2016)}]{chatterjee2016}%
  \BibitemOpen
  \bibfield  {author} {\bibinfo {author} {\bibfnamefont {S.}~\bibnamefont {Chatterjee}}\ and\ \bibinfo {author} {\bibfnamefont {S.}~\bibnamefont {Sachdev}},\ }\href {\doibase 10.1103/PhysRevB.94.205117} {\bibfield  {journal} {\bibinfo  {journal} {Phys. Rev. B}\ }\textbf {\bibinfo {volume} {94}},\ \bibinfo {pages} {205117} (\bibinfo {year} {2016})}\BibitemShut {NoStop}%
\bibitem [{\citenamefont {Zhang}\ and\ \citenamefont {Sachdev}(2020)}]{zhangyahui2020}%
  \BibitemOpen
  \bibfield  {author} {\bibinfo {author} {\bibfnamefont {Y.-H.}\ \bibnamefont {Zhang}}\ and\ \bibinfo {author} {\bibfnamefont {S.}~\bibnamefont {Sachdev}},\ }\href {\doibase 10.1103/PhysRevResearch.2.023172} {\bibfield  {journal} {\bibinfo  {journal} {Phys. Rev. Res.}\ }\textbf {\bibinfo {volume} {2}},\ \bibinfo {pages} {023172} (\bibinfo {year} {2020})}\BibitemShut {NoStop}%
\bibitem [{\citenamefont {Abrikosov}(1969)}]{abrikosov1969galvanomagnetic}%
  \BibitemOpen
  \bibfield  {author} {\bibinfo {author} {\bibfnamefont {A.}~\bibnamefont {Abrikosov}},\ }\href {http://83.149.229.155/cgi-bin/dn/e_029_04_0746.pdf} {\bibfield  {journal} {\bibinfo  {journal} {Sov. Phys. JETP}\ }\textbf {\bibinfo {volume} {29}},\ \bibinfo {pages} {746} (\bibinfo {year} {1969})}\BibitemShut {NoStop}%
\bibitem [{\citenamefont {Konik}\ \emph {et~al.}(2006)\citenamefont {Konik}, \citenamefont {Rice},\ and\ \citenamefont {Tsvelik}}]{KRT_PhysRevLett.96.086407}%
  \BibitemOpen
  \bibfield  {author} {\bibinfo {author} {\bibfnamefont {R.~M.}\ \bibnamefont {Konik}}, \bibinfo {author} {\bibfnamefont {T.~M.}\ \bibnamefont {Rice}}, \ and\ \bibinfo {author} {\bibfnamefont {A.~M.}\ \bibnamefont {Tsvelik}},\ }\href {\doibase 10.1103/PhysRevLett.96.086407} {\bibfield  {journal} {\bibinfo  {journal} {Phys. Rev. Lett.}\ }\textbf {\bibinfo {volume} {96}},\ \bibinfo {pages} {086407} (\bibinfo {year} {2006})}\BibitemShut {NoStop}%
\bibitem [{\citenamefont {Ng}(2005)}]{Ng_PhysRevB.71.172509}%
  \BibitemOpen
  \bibfield  {author} {\bibinfo {author} {\bibfnamefont {T.-K.}\ \bibnamefont {Ng}},\ }\href {\doibase 10.1103/PhysRevB.71.172509} {\bibfield  {journal} {\bibinfo  {journal} {Phys. Rev. B}\ }\textbf {\bibinfo {volume} {71}},\ \bibinfo {pages} {172509} (\bibinfo {year} {2005})}\BibitemShut {NoStop}%
\bibitem [{\citenamefont {Stanescu}\ and\ \citenamefont {Kotliar}(2006)}]{Kotliar_PhysRevB.74.125110}%
  \BibitemOpen
  \bibfield  {author} {\bibinfo {author} {\bibfnamefont {T.~D.}\ \bibnamefont {Stanescu}}\ and\ \bibinfo {author} {\bibfnamefont {G.}~\bibnamefont {Kotliar}},\ }\href {\doibase 10.1103/PhysRevB.74.125110} {\bibfield  {journal} {\bibinfo  {journal} {Phys. Rev. B}\ }\textbf {\bibinfo {volume} {74}},\ \bibinfo {pages} {125110} (\bibinfo {year} {2006})}\BibitemShut {NoStop}%
\bibitem [{\citenamefont {Kent}\ \emph {et~al.}(2008)\citenamefont {Kent}, \citenamefont {Saha-Dasgupta}, \citenamefont {Jepsen}, \citenamefont {Andersen}, \citenamefont {Macridin}, \citenamefont {Maier}, \citenamefont {Jarrell},\ and\ \citenamefont {Schulthess}}]{Kent_PhysRevB.78.035132}%
  \BibitemOpen
  \bibfield  {author} {\bibinfo {author} {\bibfnamefont {P.~R.~C.}\ \bibnamefont {Kent}}, \bibinfo {author} {\bibfnamefont {T.}~\bibnamefont {Saha-Dasgupta}}, \bibinfo {author} {\bibfnamefont {O.}~\bibnamefont {Jepsen}}, \bibinfo {author} {\bibfnamefont {O.~K.}\ \bibnamefont {Andersen}}, \bibinfo {author} {\bibfnamefont {A.}~\bibnamefont {Macridin}}, \bibinfo {author} {\bibfnamefont {T.~A.}\ \bibnamefont {Maier}}, \bibinfo {author} {\bibfnamefont {M.}~\bibnamefont {Jarrell}}, \ and\ \bibinfo {author} {\bibfnamefont {T.~C.}\ \bibnamefont {Schulthess}},\ }\href {\doibase 10.1103/PhysRevB.78.035132} {\bibfield  {journal} {\bibinfo  {journal} {Phys. Rev. B}\ }\textbf {\bibinfo {volume} {78}},\ \bibinfo {pages} {035132} (\bibinfo {year} {2008})}\BibitemShut {NoStop}%
\bibitem [{\citenamefont {Doiron-Leyraud}\ \emph {et~al.}(2007)\citenamefont {Doiron-Leyraud}, \citenamefont {Proust}, \citenamefont {LeBoeuf}, \citenamefont {Levallois}, \citenamefont {Bonnemaison}, \citenamefont {Liang}, \citenamefont {Bonn}, \citenamefont {Hardy},\ and\ \citenamefont {Taillefer}}]{doiron2007quantum}%
  \BibitemOpen
  \bibfield  {author} {\bibinfo {author} {\bibfnamefont {N.}~\bibnamefont {Doiron-Leyraud}}, \bibinfo {author} {\bibfnamefont {C.}~\bibnamefont {Proust}}, \bibinfo {author} {\bibfnamefont {D.}~\bibnamefont {LeBoeuf}}, \bibinfo {author} {\bibfnamefont {J.}~\bibnamefont {Levallois}}, \bibinfo {author} {\bibfnamefont {J.-B.}\ \bibnamefont {Bonnemaison}}, \bibinfo {author} {\bibfnamefont {R.}~\bibnamefont {Liang}}, \bibinfo {author} {\bibfnamefont {D.}~\bibnamefont {Bonn}}, \bibinfo {author} {\bibfnamefont {W.}~\bibnamefont {Hardy}}, \ and\ \bibinfo {author} {\bibfnamefont {L.}~\bibnamefont {Taillefer}},\ }\href {https://doi.org/10.1038/nature05872} {\bibfield  {journal} {\bibinfo  {journal} {Nature}\ }\textbf {\bibinfo {volume} {447}},\ \bibinfo {pages} {565} (\bibinfo {year} {2007})}\BibitemShut {NoStop}%
\bibitem [{\citenamefont {Bari{\v{s}}i{\'c}}\ \emph {et~al.}(2013)\citenamefont {Bari{\v{s}}i{\'c}}, \citenamefont {Badoux}, \citenamefont {Chan}, \citenamefont {Dorow}, \citenamefont {Tabis}, \citenamefont {Vignolle}, \citenamefont {Yu}, \citenamefont {B{\'e}ard}, \citenamefont {Zhao}, \citenamefont {Proust} \emph {et~al.}}]{barivsic2013universal}%
  \BibitemOpen
  \bibfield  {author} {\bibinfo {author} {\bibfnamefont {N.}~\bibnamefont {Bari{\v{s}}i{\'c}}}, \bibinfo {author} {\bibfnamefont {S.}~\bibnamefont {Badoux}}, \bibinfo {author} {\bibfnamefont {M.~K.}\ \bibnamefont {Chan}}, \bibinfo {author} {\bibfnamefont {C.}~\bibnamefont {Dorow}}, \bibinfo {author} {\bibfnamefont {W.}~\bibnamefont {Tabis}}, \bibinfo {author} {\bibfnamefont {B.}~\bibnamefont {Vignolle}}, \bibinfo {author} {\bibfnamefont {G.}~\bibnamefont {Yu}}, \bibinfo {author} {\bibfnamefont {J.}~\bibnamefont {B{\'e}ard}}, \bibinfo {author} {\bibfnamefont {X.}~\bibnamefont {Zhao}}, \bibinfo {author} {\bibfnamefont {C.}~\bibnamefont {Proust}},  \emph {et~al.},\ }\href {https://doi.org/10.1038/nphys2792} {\bibfield  {journal} {\bibinfo  {journal} {Nature Physics}\ }\textbf {\bibinfo {volume} {9}},\ \bibinfo {pages} {761} (\bibinfo {year} {2013})}\BibitemShut {NoStop}%
\bibitem [{\citenamefont {Allais}\ \emph {et~al.}(2014)\citenamefont {Allais}, \citenamefont {Chowdhury},\ and\ \citenamefont {Sachdev}}]{allais2014connecting}%
  \BibitemOpen
  \bibfield  {author} {\bibinfo {author} {\bibfnamefont {A.}~\bibnamefont {Allais}}, \bibinfo {author} {\bibfnamefont {D.}~\bibnamefont {Chowdhury}}, \ and\ \bibinfo {author} {\bibfnamefont {S.}~\bibnamefont {Sachdev}},\ }\href {https://doi.org/10.1038/ncomms6771} {\bibfield  {journal} {\bibinfo  {journal} {Nature communications}\ }\textbf {\bibinfo {volume} {5}},\ \bibinfo {pages} {5771} (\bibinfo {year} {2014})}\BibitemShut {NoStop}%
\bibitem [{\citenamefont {Bonetti}\ \emph {et~al.}(2024)\citenamefont {Bonetti}, \citenamefont {Christos},\ and\ \citenamefont {Sachdev}}]{bonetti2024quantum}%
  \BibitemOpen
  \bibfield  {author} {\bibinfo {author} {\bibfnamefont {P.~M.}\ \bibnamefont {Bonetti}}, \bibinfo {author} {\bibfnamefont {M.}~\bibnamefont {Christos}}, \ and\ \bibinfo {author} {\bibfnamefont {S.}~\bibnamefont {Sachdev}},\ }\href {\doibase 10.1073/pnas.2418633121} {\bibfield  {journal} {\bibinfo  {journal} {Proceedings of the National Academy of Sciences}\ }\textbf {\bibinfo {volume} {121}},\ \bibinfo {pages} {e2418633121} (\bibinfo {year} {2024})}\BibitemShut {NoStop}%
\bibitem [{\citenamefont {Ashcroft}\ and\ \citenamefont {Mermin}(1976)}]{ashcroft1976solid}%
  \BibitemOpen
  \bibfield  {author} {\bibinfo {author} {\bibfnamefont {N.}~\bibnamefont {Ashcroft}}\ and\ \bibinfo {author} {\bibfnamefont {N.}~\bibnamefont {Mermin}},\ }\href@noop {} {\emph {\bibinfo {title} {Solid State Physics}}}\ (\bibinfo  {publisher} {England: Thomson},\ \bibinfo {year} {1976})\BibitemShut {NoStop}%
\bibitem [{\citenamefont {Abrikosov}(2017)}]{abrikosov2017fundamentals}%
  \BibitemOpen
  \bibfield  {author} {\bibinfo {author} {\bibfnamefont {A.}~\bibnamefont {Abrikosov}},\ }\href@noop {} {\emph {\bibinfo {title} {Fundamentals of the Theory of Metals}}}\ (\bibinfo  {publisher} {Courier Dover Publications},\ \bibinfo {year} {2017})\BibitemShut {NoStop}%
\bibitem [{\citenamefont {Chan}\ \emph {et~al.}(2020)\citenamefont {Chan}, \citenamefont {McDonald}, \citenamefont {Ramshaw}, \citenamefont {Betts}, \citenamefont {Shekhter}, \citenamefont {Bauer},\ and\ \citenamefont {Harrison}}]{chan_PNAS}%
  \BibitemOpen
  \bibfield  {author} {\bibinfo {author} {\bibfnamefont {M.~K.}\ \bibnamefont {Chan}}, \bibinfo {author} {\bibfnamefont {R.~D.}\ \bibnamefont {McDonald}}, \bibinfo {author} {\bibfnamefont {B.~J.}\ \bibnamefont {Ramshaw}}, \bibinfo {author} {\bibfnamefont {J.~B.}\ \bibnamefont {Betts}}, \bibinfo {author} {\bibfnamefont {A.}~\bibnamefont {Shekhter}}, \bibinfo {author} {\bibfnamefont {E.~D.}\ \bibnamefont {Bauer}}, \ and\ \bibinfo {author} {\bibfnamefont {N.}~\bibnamefont {Harrison}},\ }\href {\doibase 10.1073/pnas.1914166117} {\bibfield  {journal} {\bibinfo  {journal} {Proceedings of the National Academy of Sciences}\ }\textbf {\bibinfo {volume} {117}},\ \bibinfo {pages} {9782} (\bibinfo {year} {2020})}\BibitemShut {NoStop}%
\bibitem [{\citenamefont {Gerber}\ \emph {et~al.}(2015)\citenamefont {Gerber}, \citenamefont {Jang}, \citenamefont {Nojiri}, \citenamefont {Matsuzawa}, \citenamefont {Yasumura}, \citenamefont {Bonn}, \citenamefont {Liang}, \citenamefont {Hardy}, \citenamefont {Islam}, \citenamefont {Mehta}, \citenamefont {Song}, \citenamefont {Sikorski}, \citenamefont {Stefanescu}, \citenamefont {Feng}, \citenamefont {Kivelson}, \citenamefont {Devereaux}, \citenamefont {Shen}, \citenamefont {Kao}, \citenamefont {Lee}, \citenamefont {Zhu},\ and\ \citenamefont {Lee}}]{YBCO}%
  \BibitemOpen
  \bibfield  {author} {\bibinfo {author} {\bibfnamefont {S.}~\bibnamefont {Gerber}}, \bibinfo {author} {\bibfnamefont {H.}~\bibnamefont {Jang}}, \bibinfo {author} {\bibfnamefont {H.}~\bibnamefont {Nojiri}}, \bibinfo {author} {\bibfnamefont {S.}~\bibnamefont {Matsuzawa}}, \bibinfo {author} {\bibfnamefont {H.}~\bibnamefont {Yasumura}}, \bibinfo {author} {\bibfnamefont {D.~A.}\ \bibnamefont {Bonn}}, \bibinfo {author} {\bibfnamefont {R.}~\bibnamefont {Liang}}, \bibinfo {author} {\bibfnamefont {W.~N.}\ \bibnamefont {Hardy}}, \bibinfo {author} {\bibfnamefont {Z.}~\bibnamefont {Islam}}, \bibinfo {author} {\bibfnamefont {A.}~\bibnamefont {Mehta}}, \bibinfo {author} {\bibfnamefont {S.}~\bibnamefont {Song}}, \bibinfo {author} {\bibfnamefont {M.}~\bibnamefont {Sikorski}}, \bibinfo {author} {\bibfnamefont {D.}~\bibnamefont {Stefanescu}}, \bibinfo {author} {\bibfnamefont {Y.}~\bibnamefont {Feng}}, \bibinfo {author} {\bibfnamefont {S.~A.}\ \bibnamefont {Kivelson}}, \bibinfo {author} {\bibfnamefont {T.~P.}\ \bibnamefont
  {Devereaux}}, \bibinfo {author} {\bibfnamefont {Z.-X.}\ \bibnamefont {Shen}}, \bibinfo {author} {\bibfnamefont {C.-C.}\ \bibnamefont {Kao}}, \bibinfo {author} {\bibfnamefont {W.-S.}\ \bibnamefont {Lee}}, \bibinfo {author} {\bibfnamefont {D.}~\bibnamefont {Zhu}}, \ and\ \bibinfo {author} {\bibfnamefont {J.-S.}\ \bibnamefont {Lee}},\ }\href {\doibase 10.1126/science.aac6257} {\bibfield  {journal} {\bibinfo  {journal} {Science}\ }\textbf {\bibinfo {volume} {350}},\ \bibinfo {pages} {949} (\bibinfo {year} {2015})}\BibitemShut {NoStop}%
\end{thebibliography}%

\end{document}